\documentclass[namedreferences]{solarphysics}

\usepackage[hyperref,optionalrh,showbiblabels]{spr-sola-addons} % For Solar Physics 
\usepackage{graphicx}        % For eps figures, newer & more powerfull
\usepackage{color}           % For color text: \color command
\usepackage{breakurl}        % For breaking URLs easily trough lines
\newcommand{\eg}{{\it e.g.}}
\newcommand{\ie}{{\it i.e.}}

\begin{document}

\begin{article}
\tracingmacros=2
\begin{opening}

\title{Mass Loss Evolution in the EUV Low Corona from SDO/AIA Data\\ {\it Solar Physics}}

\author[addressref={aff1},corref, email={flopez@icate-conicet.gob.ar}]{\inits{F.M.}\fnm{Fernando M.}~\lnm{L\'opez}} %\sep
\author[addressref={aff2},email={hebe.cremades@frm.utn.edu.ar}]{\inits{M.H.}\fnm{Hebe}~\lnm{Cremades}}%\sep
\author[addressref={aff3},email={federico@iafe.uba.ar}]{\inits{F.A.}\fnm{Federico A.}~\lnm{Nuevo}}%\sep
\author[addressref={aff1,aff4},email={lbalmaceda@icate-conicet.gob.ar}]{\inits{L.A.}\fnm{Laura A.}~\lnm{Balmaceda}}%\sep
\author[addressref={aff3,aff5},email={albert@iafe.uba.ar}]{\inits{A.M.}\fnm{Alberto M.}~\lnm{V\'asquez}}%\sep

\address[id=aff1]{Instituto de Ciencias Astron\'omicas, de la Tierra y del Espacio (ICATE), CONICET, San Juan, Argentina.}
\address[id=aff2]{Universidad Tecnol\'ogica Nacional--Facultad Regional Mendoza, CONICET, CEDS, Mendoza, Argentina.}
\address[id=aff3]{Instituto de Astronom\'ia y F\'isica del Espacio (IAFE, CONICET-UBA), Buenos Aires, Argentina.}
\address[id=aff4]{Instituto Nacional de Pesquisas Espaciais (INPE), S\~{a}o Jos\'e dos Campos, Brazil.}
\address[id=aff5]{Universidad Nacional de Tres de Febrero (UNTREF), Ingenier\'ia Ambiental, Departamento de Ciencia y Tecnolog\'ia, Buenos Aires, Argentina.}

\runningauthor{L\'opez et al.}
\runningtitle{Mass Loss in the EUV Low Corona}

\begin{abstract}
We carry out an analysis of the evacuated mass from three coronal dimming regions observed by the {\it Atmospheric Imaging Assembly} (AIA) on board the {\it Solar Dynamics Observatory}. The three events are unambiguously identified with white-light coronal mass ejections (CMEs) associated in turn with surface activity of diverse nature: an impulsive (M-class) flare, a weak (B-class) flare and a filament eruption without a flare. The use of three AIA coronal passbands allows applying a differential emission measure technique to define the dimming regions and identify their evacuated mass through the analysis of the electronic density depletion associated to the eruptions. The temporal evolution of the mass loss from the three dimmings can be approximated by an exponential equation followed by a linear fit. We determine the mass of the associated CMEs from COR2 data. The results show that the evacuated masses from the low corona represent a considerable amount of the mass of the CMEs. We also find that plasma is still being evacuated from the low corona at the time when the CMEs reach the COR2 field of view. The temporal evolution of the angular width of  the CMEs, of the dimming regions in the low corona, and of the flux registered by GOES in soft X-rays are all in close relation with the behavior of mass evacuation from the low corona. We discuss the implications of our findings toward a better understanding of the temporal evolution of several parameters associated to the analyzed dimmings and CMEs.
\end{abstract}
\end{opening} 
%-------------------------------------------------
\tracingmacros=0

\section{Introduction}
     \label{S-Introduction} 

Coronal mass ejections (CMEs) are one of the most energetic events generated by the Sun. They transport millions of tons of plasma from the corona into interplanetary space, where they can generate geomagnetic storms on Earth affecting human activities. This is the main reason for which the CME phenomenon has been actively investigated during the last four decades.

The first studies about CMEs showed that they are generally associated with filament eruptions and flares \citep{Munro1979}. However, the origin and nature of the bulk of the CME mass was unclear. The use of Skylab coronagraph data allowed \citet{Hildner1975} to speculate that the leading edge of mass ejections is mainly composed of material from the lower corona. The analysis of the low-corona observations obtained with a soft X-ray telescope aboard Skylab made possible the discovery of intensity depletions at the location of expanding arcs for an event observed on the solar limb \citep{Rust1976}. Later, {\it Yohkoh} soft X-ray images allowed the observation of intensity depletions associated with halo CMEs for events near solar center \citep{Hudson1998, Sterling1997}. Similar depletions of intensity were detected in EUV coronal wavelengths \citep{Thompson1998,Zarro1999} using data from the {\it Extreme-ultraviolet Imaging Telescope} \citep[EIT:][]{Delaboudiniere1995}. These depletions were similar in appearance to coronal holes, reason for which they were referred to as ``transient coronal holes'' \citep{Zhukov2004}. Further studies confirmed the spatial and temporal association of the onset of CMEs with these transient coronal holes \citep[\eg,][]{Harrison2003}, nowadays known as coronal dimmings.

Using EUV spectroscopic observations from the {\it Coronal Diagnostic Spectrometer} \citep[CDS:][]{Harrison1995}, \citet{Harrison2000} and \citet{Harrison2003} determined the mass loss from dimmings observed on the solar limb. They found that the mass loss in the dimming regions was of the same order of magnitude of the values estimated for the masses of their associated CMEs. In a later study, \citet{Aschwanden2009} used data from the {\it Extreme-Ultraviolet Imager} \citep[EUVI:][]{Wuelser2004} on board the {\it Solar-Terrestrial Relations Observatory} (STEREO) to obtain the mass loss of 8 dimming events, most of them observed on the limb. They compared the results with the mass of the corresponding CMEs determined from the STEREO COR2 coronagraphs, and found a good agreement. In a more recent study, using data from the {\it EUV Imaging Spectrometer} on board {\it Hinode}, \citet{Tian2012} found that the mass loss from three dimming regions was in the range of 20\% - 60\% of the masses of their associated CMEs. 

It is widely accepted that coronal dimming regions are related to depletions of plasma density in areas associated to eruptions. Dimmings are more easily noticeable at the 195\,\AA\ and 211\,\AA\, wavelengths rather than, for instance, at lower-temperature wavelengths such as at 171\,\AA. This is probably due to the fact that most of the cooler plasma is not ejected during the opening of loops, remaining gravitationally bound \citep{Robbrecht2010}. 

Nowadays, most of the space wheather forecasts rely on the fact that an Earthward CME is detected either by the {\it Large Angle Spectrometric Coronagraph} \citep[LASCO:][]{Brueckner1995} onboard the {\it Solar and Heliospheric Observatory} (SOHO), or by the coronagraphs onboard the STEREO spacecraft. Given the imminent conclusion of the SOHO mission, the strong position dependence of the STEREO remote-sensing data, and the loss of one of the STEREO spacecraft (see \url{http://stereo.gsfc.nasa.gov/status.shtml}), the determination of CME masses through low-coronal dimming observations would provide important information regarding  potentially geoeffective events. This is particularly valuable for front-sided events, given that measurements of CMEs masses are not accurate when the CME is propagating far from the plane of the sky (POS), according to the Thomson-scattering theory \citep{Billings1966}.

The dimmings analyzed in this work are associated with eruption of material from the low corona that can be related to CMEs observed in white light. Our study represents the first analysis of the temporal evolution of mass loss in dimming regions applying a differential emission measure (DEM) technique on images obtained by the {\it Atmospheric Imaging Assembly} \citep[AIA:][]{Lemen2012} on board the {\it Solar Dynamics Observatory} (SDO). The deduced values are compared with the masses of the associated CMEs, carefully determined from nearly-quadrature coronagraphic images. In the present paper, we will use ``mass loss'' and ``evacuated mass'' indistinctly to refer to the mass difference in a dimming region at a time before and after the eruption.

The applied DEM technique makes use of data taken by three narrowband EUV imagers sensitive to a wide range of temperatures, thus mitigating plasma temperature effects. Other processes that can produce a decrease in the intensity registered at a single wavelength band by EUV telescopes include Doppler effects \citep{Mason2014}. However, these mechanisms (both Doppler dimming and passband shift) do not play a significant role in the present analysis, which is based on the narrowband EUV images taken by the AIA instrument.

In the next section we describe the dimming events under study, their associated CMEs and the data used for the analysis. In section \ref{S-Evacuated-Mass} we provide a description of the DEM technique, followed by the method we use to estimate the EUV mass loss from the low corona. In section \ref{S-CME_Mass} we perform the determination of the CMEs' masses from white-light images, while in Section \ref{S-Mass-Comparison} we compare the EUV and white-light results. Section \ref{S-Xray} addresses the temporal relationship between mass loss and X-ray flux profiles. Finally, in Section \ref{S-Summary} we summarize our results and present the conclusions.

\section{Selected Events} %%%%%%%%%%%%%%
  \label{S-Events}
  
We present the analysis of three events characterized by the presence of an EUV dimming and an associated CME. All eruptions have been observed in the EUV by AIA on board the SDO. AIA provides nearly-simultaneous full-disk images of the solar corona and transition region in multiple wavelengths with unprecedented temporal and spatial resolution. Full-disk AIA images are recorded with a 12-second cadence and a size of $4096^2$ pixels, resulting in a spatial resolution of 1.5 arcsec. 

The associated CMEs were observed in white light by the COR1 and COR2 coronagraphs, wich are part of the {\it Sun-Earth Connection Coronal and Heliospheric Investigation} \citep[SECCHI:][]{Howard2008} on board the STEREO mission. The COR1 coronagraph observes the white-light corona with a FOV of 1.4\,--\,4.0 R$_{\odot}$, with a 5-minute cadence. The COR2 instrument has a FOV of 2.5\,--\,15 R$_{\odot}$ and images the white-light corona with a spatial resolution of 30 arcsec. 

The events under study took place between May and November 2010, when the positions of the STEREO twin spacecraft formed an angle of $\approx$\,70$^{\circ}$ to 85$^{\circ}$ with the Sun--Earth line. To avoid limb events from the Earth's perspective, the location of the eruptions originating the dimmings were required to take place in the range of [-50$^{\circ}$,50$^{\circ}$] heliographic longitude. At the same time, this implies that their associated CMEs propagate with an angle close to the POS for at least one of the STEREO spacecraft.

The first dimming event (Event 1) occurred on 23 May 2010, starting at \mbox{16:52 UT} as a B1.4 flare with heliographic coordinates N19 W12, according to the XRT Flare Catalog \citep[][\url{http://xrt.cfa.harvard.edu/flare_catalog/}]{Watanabe2012}. This flare was not registered by the {\it Geostationary Operational Environmental Satellites} (GOES). For this reason, we perform an analysis of the X-ray and EUV data to verify the existence of the flare. The X-ray and EUV data show the presence of a flare at the location the eruption, supporting the report by the XRT Flare Catalog. EUV images of the low corona show a quiescent filament in eruption from these coordinates, far from any active region. The associated CME is observed close to the POS of the COR2-A and COR2-B, arising in the field of view (FOV) of these instruments at 17:39 UT and 17:54 UT respectively.

The second dimming (Event 2) took place on 7 August 2010. This event started with a GOES M1.0 flare at 17:55 UT. The event occurred in active region (AR) NOAA 11093, with GOES flare coordinates N14 E37. This dimming expanded in association with a coronal wave event, while its corresponding CME was observed by SOHO/LASCO and COR2-B as a partial halo CME. It emerged in the FOV of COR2-A at 18:39 UT, where it propagates close to the POS. \citet{Mason2014} investigated this event, and although they did not estimate the mass evacuated from the dimming, from the analysis of light curves for different wavelengths from AIA images and EVE data they conclude that nearly 100\% of the dimming was due to mass loss in the corona.

The third dimming event (Event 3) was observed by AIA on 30 November 2010. It began as a filament eruption starting at around 17:35 UT with heliographic coordinates N13 E32. There was no solar flare registered by GOES or the XRT Flare Catalog which could be potentially associated to this event. For the previous events we approximate the start time and locations of the eruptions by those reported for the associated flares. However, in this case the start time and the coordinates of the eruption were derived from AIA images. The CME was observed to propagate close to the POS of COR2-A and COR2-B after 19:24 UT and 20:10 UT respectively.

\section{Determination of Evacuated Mass from the Low Corona} %%%%%%%%%%%%%%%%%%%%%%%%%%%%%%%%%%%%%%%%
 \label{S-Evacuated-Mass}      

\subsection{The Parametric DEM Technique} %%%%%%%%%%%%%%
  \label{S-dem}
  
The DEM technique makes possible the determination of density and temperature of the plasma in the low corona, by providing the thermal distribution of the plasma contained in a column along the line of sight (LOS) associated with a pixel. In this work we use the parametric DEM technique developed by \citet{Nuevo2015}, adapted for the specific case of high temporal and spatial resolution required to analyze dimmings. While DEM distributions can be determined through Markov Chain Monte Carlo (MCMC) methods \citep{Kashyap1998} or regularized inversion techniques \citep{Hannah2012} when high resolution spectra are used as input data, parametric DEM studies are a useful approach when based on narrow-band images, as in this work. Similar techniques to the one described below have been recently used, for example, to study ARs \citep{Aschwanden2011,DelZanna2013,Plowman2013}. A validation study of the parametric technique here used, which compares results to MCMC methods, is included in \citet{Nuevo2015}.

To obtain the DEM maps of the dimming regions, from which the mass loss is estimated, we use the AIA coronal passbands centered at 171\,\AA\ (Fe {\sc ix}), 193\,\AA\ (Fe {\sc xii}/Fe {\sc xxiv}) and 211\,\AA\ (Fe {\sc xiv}). For the three aforementioned AIA bands the temperature response function peaks at $\approx$~0.9, 1.5, and 1.9 MK, respectively, covering the temperature range characteristic of quiet Sun regions. For each passband, only three images per minute are averaged to obtain one mean image per minute, so as to increase the signal-to-noise (S/N) ratio. At this high cadence still ensure that there are no significant changes in the dimming regions. The standard calibration was applied to the AIA images while rebinning them to a size of $1024^2$ pixels, to further increase the S/N ratio and to reduce the calculation time of the DEM algorithms. In order to account for changes in the coronal structures due to differential rotation of the Sun, all images were de-rotated to a pre-event time. A $2\times2$ mean filter was finally applied to remove high-frequency noise in the images.     

The intensity $I_{k,i}$ registered by the instrument AIA in the passband $k$ in a certain pixel $i$ is given by Equation \ref{E-intensity}, where $\psi_{i}(T)$ is the DEM distribution for the pixel $i$ and $Q_k(T)$ is the temperature response function (TRF) of the respective band $k$. 

\begin{equation} \label{E-intensity}
I_{k,i} = \int Q_{k}(T)\,\psi_{i}(T)\, dT
\end{equation}

The TRF for each band is computed based on the CHIANTI atomic database and plasma emissivity model \citep{Dere1997} and using the instrumental passbands. Knowing the TRF, the challenge consists in finding for each pixel the DEM that accurately predicts the observed intensity in all three bands. The solution is implemented by considering a family of functions $\psi_{i} = F(\gamma_{i}, T)$, with parameters $\gamma_i$. When using three EUV bands, an appropriate parametric model $F(\gamma,T)$ is a single Gaussian distribution with three free parameters: the centroid T$_{0}$, the standard desviation $\sigma_T$ and the amplitude A. For each pixel $i$, the solution is obtained by finding the values of the parameters $\gamma_i$ which minimize the quadratic difference between (i) the intensity predicted by the DEM Gaussian model given by Equation \ref{E-intensity} and (ii) the observed intensity, summed over all three AIA passbands. For a detailed explanation of the method, we refer the reader to the articles by \citet{Frazin2009} and \citet{Nuevo2015}, where the DEM parametric technique is developed in the context of EUV tomography. A review on the subject has been recently published by \citep{Vasquez2016}.

Once the DEM is found, we can derive the emission measure (EM) and the temperature averaged along the LOS, by taking the zeroth and first order moments of the DEM, specifically:
\begin{eqnarray}
 EM &=& \int  \psi_i(T) \, dT\,  \\
 \langle T \rangle &=& \int   \psi_i(T) \, T/EM dT\,,
\label{Tm} 
\end{eqnarray}
while EM is the square density averaged along the LOS:
\begin{equation}\label{EMdef}
 EM = \int_0^L N_e^2(z) \, dz \,,
\end{equation}
where $z$ is the coordinate along the LOS and $L$ is the length of the plasma column.

\subsection{Definition of the Dimming Region} %%%%%%%%%%%%%%
  \label{S-dimming}
  
\begin{figure}    %%%%%%%%%%%%%%%%%% FIGURE 1
                                % includes the three top panels 
   \centerline{\hspace*{0.00\textwidth}
               \includegraphics[width=0.39\textwidth,clip=]{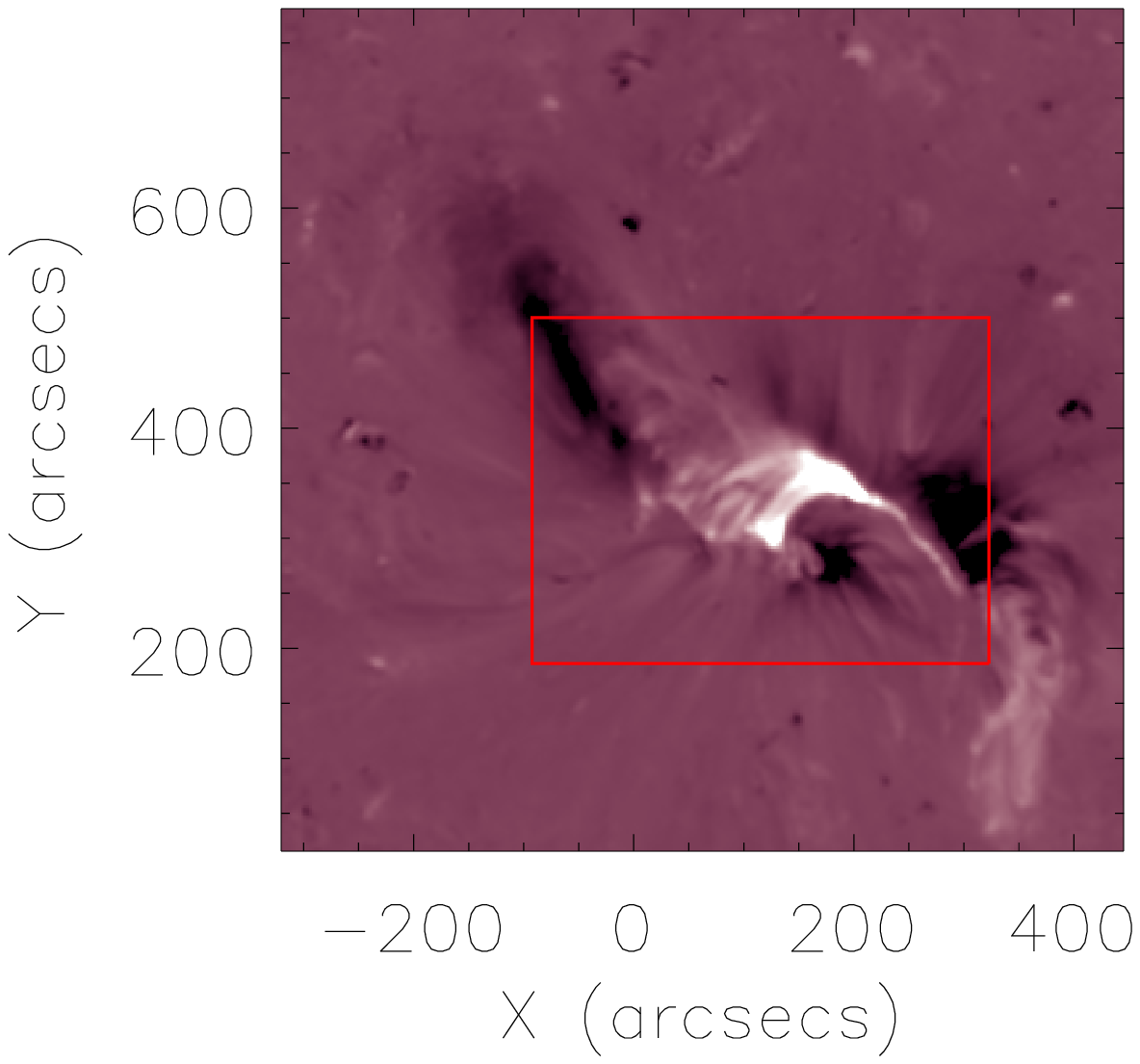}
               \hspace*{-0.01\textwidth}
               \includegraphics[width=0.502\textwidth,clip=]{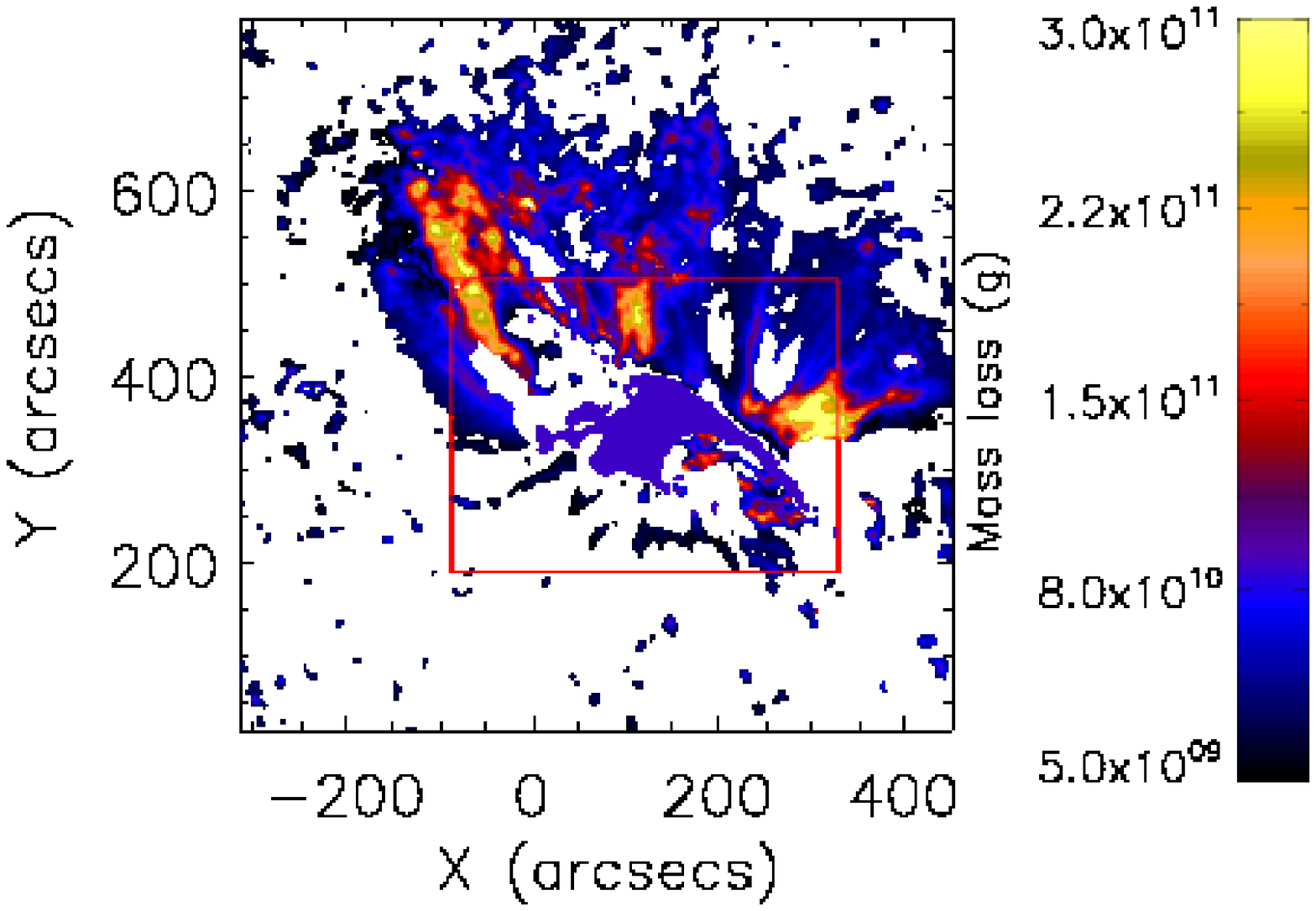}
               }
               \vspace{-0.35\textwidth}   % Shift close to the panel top 
     \centerline{\large \bf     % Includes the labels (here needs the color 
                                %   package, see beginning of this file)
      \hspace{0.05 \textwidth}  \color{black}{(a)}
      \hspace{0.33\textwidth}  \color{black}{(b)}
         \hfill}
     \vspace{0.35\textwidth}    % Shift back to the panel bottom       
  
%------------------
   \centerline{\hspace*{0.00\textwidth}
               \includegraphics[width=0.39\textwidth,clip=]{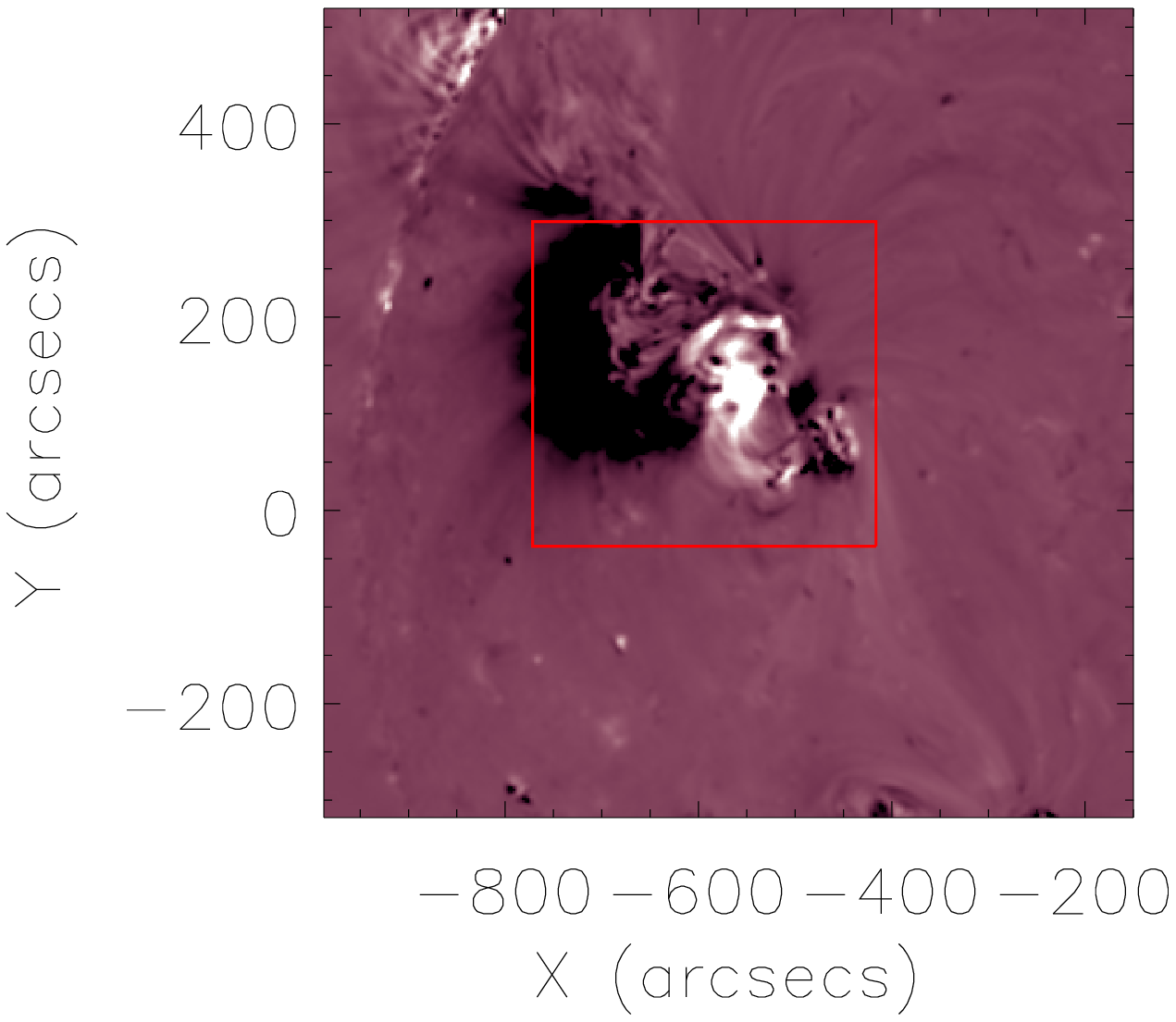}
               \hspace*{-0.01\textwidth}
               \includegraphics[width=0.502\textwidth,clip=]{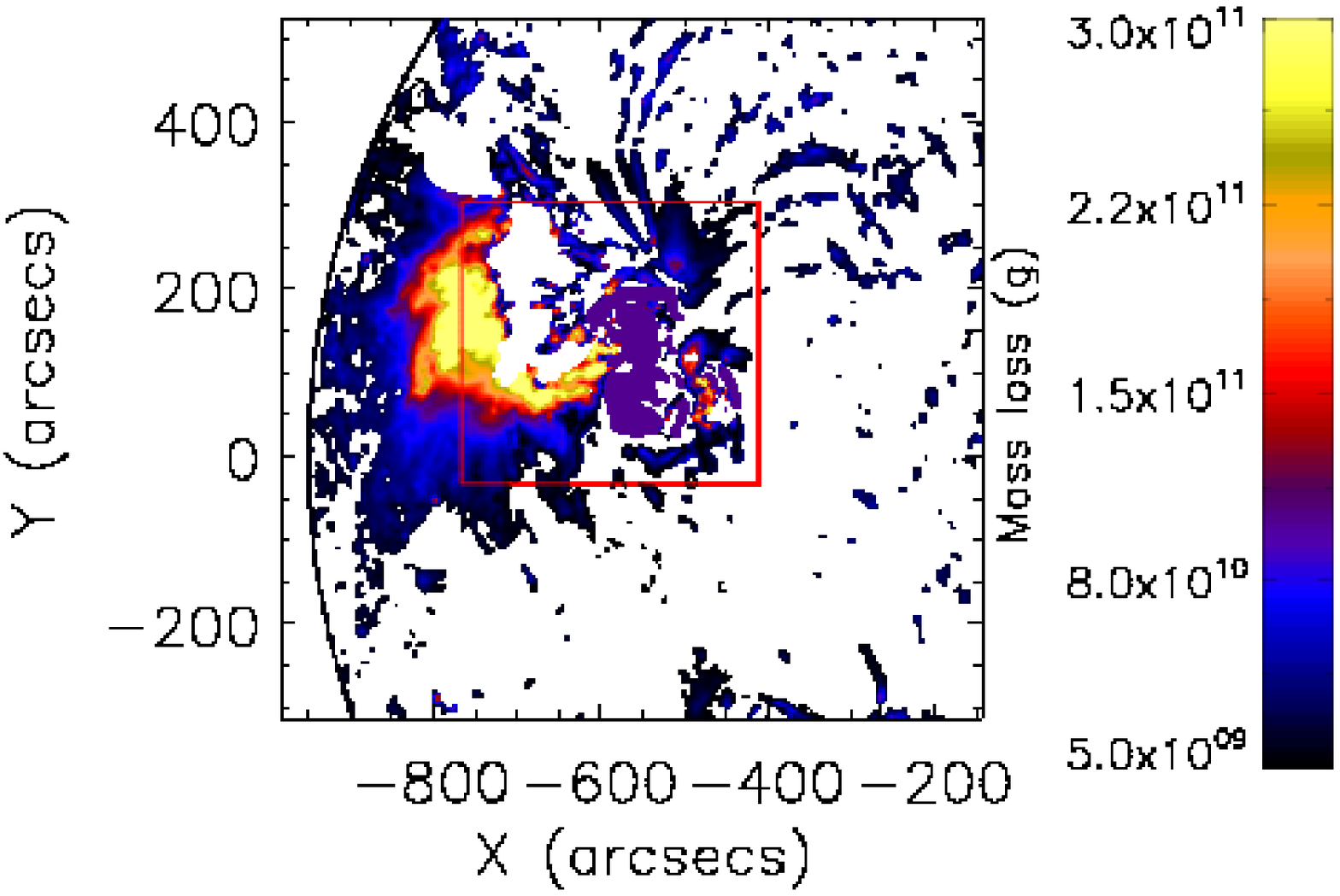}
              }
     \vspace{-0.35\textwidth}   % Shift close to the panel top 
     \centerline{\large \bf     % Includes the labels (here needs the color package)
      \hspace{0.05\textwidth} \color{black}{(c)}
      \hspace{0.33\textwidth}  \color{black}{(d)}
         \hfill}
     \vspace{0.35\textwidth}    % Shift back to the panel bottom 
%------------------------     
     
     \centerline{\hspace*{0.00\textwidth}
     \includegraphics[width=0.39\textwidth,clip=]{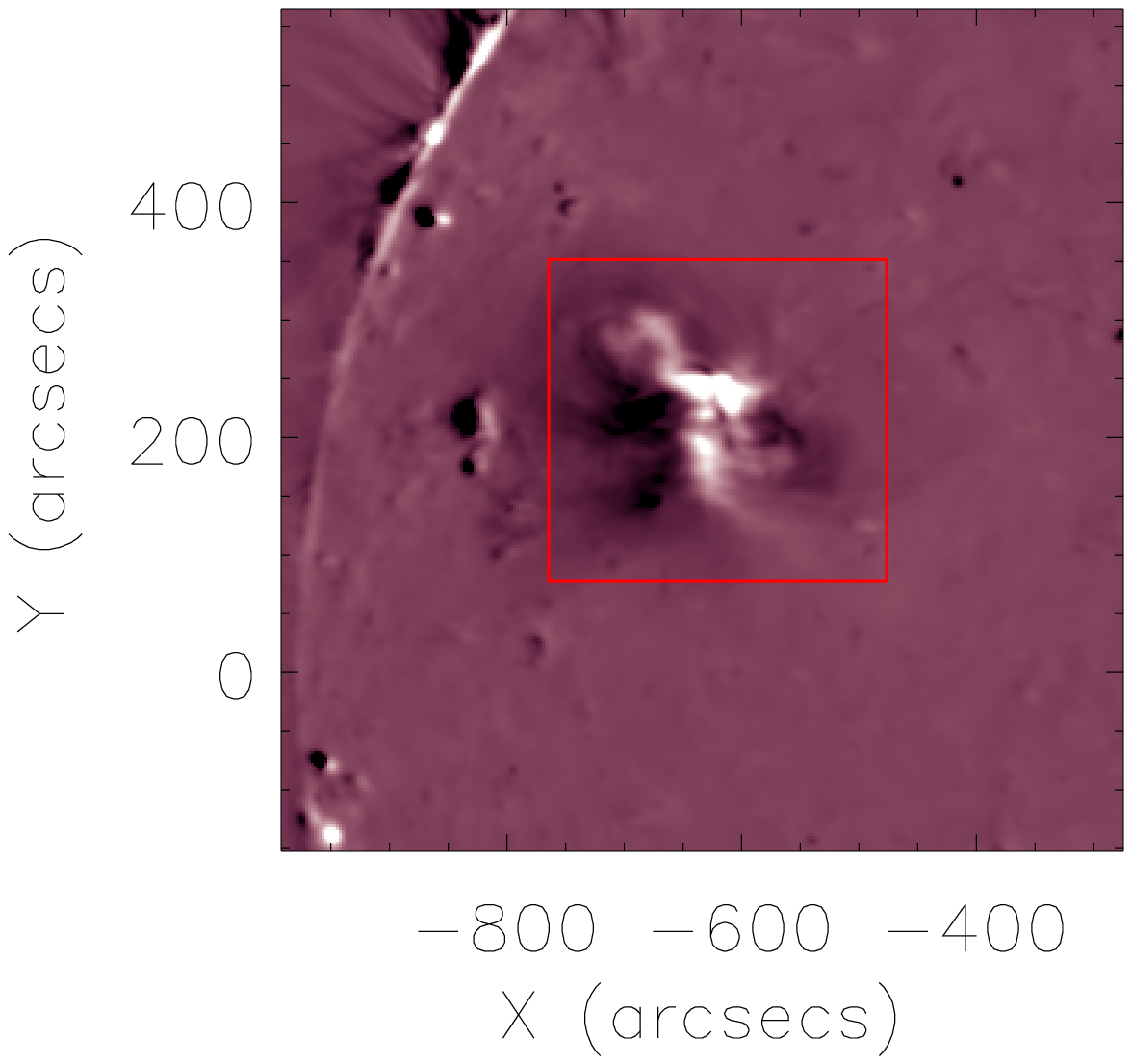}
     \hspace*{-0.01\textwidth}
     \includegraphics[width=0.502\textwidth,clip=]{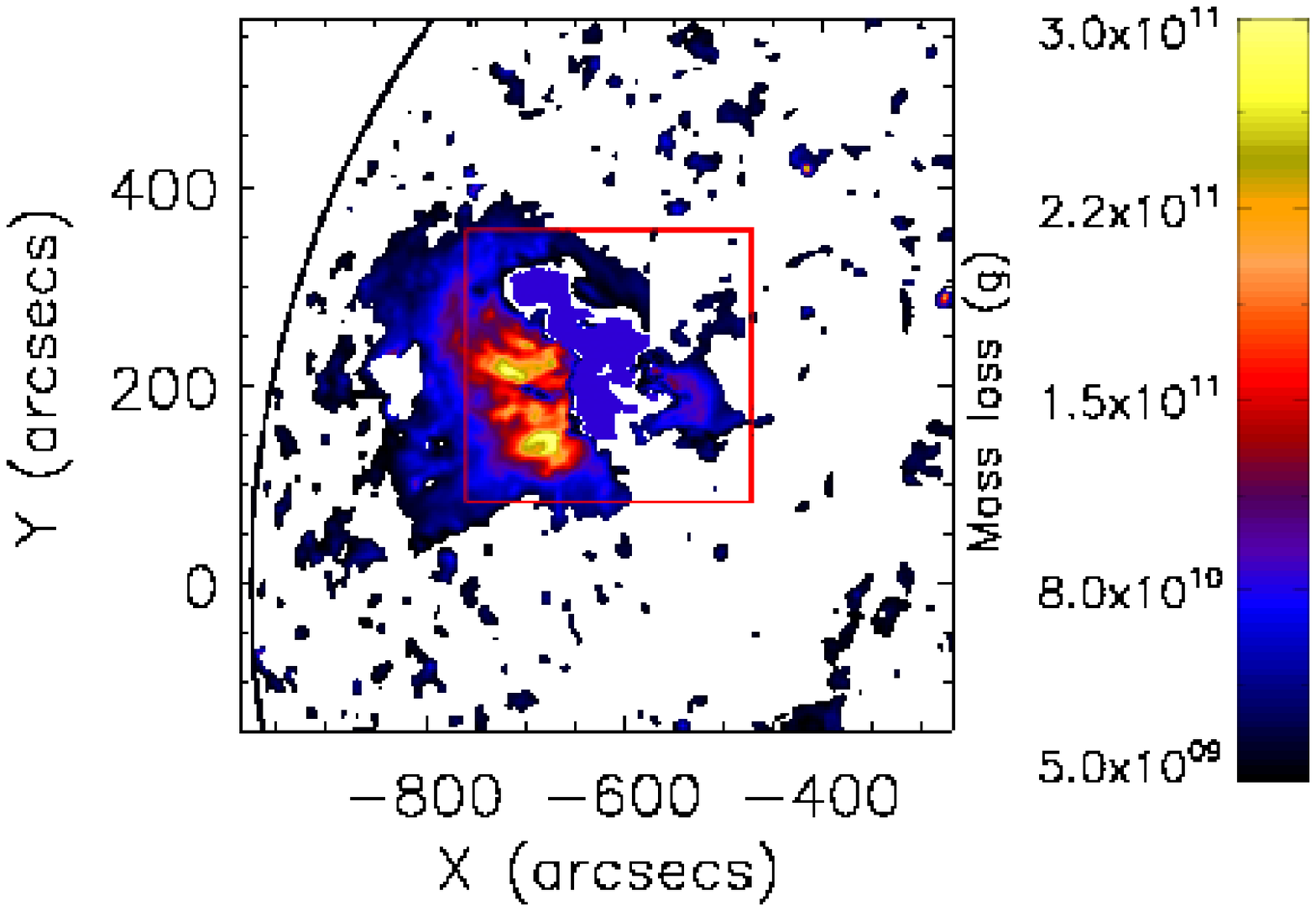}
              }
               \vspace{-0.35\textwidth}   % Shift close to the panel top 
     \centerline{\large \bf     % Includes the labels (here needs the color package)
      \hspace{0.05\textwidth} \color{black}{(e)}
      \hspace{0.33\textwidth}  \color{black}{(f)}
         \hfill}
     \vspace{0.35\textwidth}    % Shift back to the panel bottom 
\caption{ROI for the events under study. The upper row corresponds to the 23 May 2010 event at 19:00 UT, the middle row to the 7 August 2010 at 21:00 UT, and the lower row to the 30 November 2010 at 20:00 UT. Left column: 211 \AA\ base-difference AIA images showing dimmings (dark areas) and post-eruptive loops (bright features). Right column: maps of evacuated mass. The color code indicates mass loss in grams. The coordinates in all plots are given in heliocentric-cartesian coordinates.}
   \label{Fig1}
   \end{figure}

The usual criterion to identify a dimming in EUV images, assumed to be caused by the evacuation of plasma in an eruption region, is to detect an intensity decrease by using a single coronal passband, for  example, 171\,\AA\ \citep[\eg][]{Alipour2012}, 195\,\AA\ \citep[\eg][]{Attrill2010, Reinard2008}, 193\,\AA\ \citep[\eg][]{Krista2013}. However, a dimming may exhibit different characteristics depending on the observation wavelength. To avoid an arbitrary selection of the dimming region by considering any particular AIA passband, we use EM maps constructed from the combination of the three passbands. As the EM is proportional to the square of the local density of plasma, it is a better indicator for the identification of the dimming. Instead of using the full image, we first select a region of interest (ROI) to determine the dimming areas of the events under study. The AIA 211\,\AA\ base-difference ROIs for the three events under study are presented in the left panels of Figure \ref{Fig1}. In these panels, the dark areas denote the coronal dimmings while the bright regions correspond to the post-eruptive loops.

The EM is obtained for all pixels in the ROI, both from pre-eruption and post-eruption images. For each pixel, the difference $\Delta EM$ between the post-eruption value ($EM^{pos}$) and the pre-eruption value ($EM^{pre}$) is computed. A pixel $i$ is considered to belong to the dimming region if the following condition is met,
\begin{equation}\label{eq-condition}
 \Delta EM_{i} = (EM^{pos}_{i} - EM^{pre}_{i} ) \leq \delta  
\end{equation}

\noindent where $\delta$ is the median value of $\Delta EM$ for all pixels $i$ that meet $\Delta EM_i\leq 0$.

The right column in Figure \ref{Fig1} shows color-coded maps of evacuated mass $\Delta M_i$ (computed through Equation \ref{Eq-deltaM} explained below) for all pixels $i$ that fullfill the condition in Equation \ref{eq-condition}. The deepest intensity drops in the difference images of the left column agree with those regions where the evacuated mass is higher in the right-column color maps, as will be ascertained in the following section.

One of the main difficulties in the determination of the evacuated mass is due to the bright post-eruptive loops that appear after the eruption of a CME and cover part of the dimming region. These bright loops contribute significantly to the emission detected in the coronal passbands of AIA, resulting in an increase of the density obtained by the DEM analysis. Furthermore, these regions of bright post-eruptive loops, which cover a significant area of the eruption site and are part of the eruptive process, are plausibly a significant source of mass evacuation. To overcome this problem, we opt for replacing the values in the pixels belonging to the post-eruptive loops by a more representative value, determined as follows. We first consider a box that encloses the post-eruptive loops (red box in Figure \ref{Fig1}). Next, in the EM base-difference maps, we select all pixels $i$ for which $\Delta EM_i \geq \mu+\sigma$, where $\mu$ and $\sigma$ are the mean and the standard deviation of $\Delta EM$, respectively, for all pixels inside the red box.

Once the pixels containing post-eruptive loops are found, we consider that each of these pixels contributes with a loss of mass equal to the mean value given by the pixels inside the box that fullfill the condition given by Equation \ref{eq-condition}. Given that it is expected that a large --or even the largest-- fraction of the erupted plasma arises from the locations below the post-eruptive loops, by assigning this mean value of evacuated mass we are most likely underestimating the contribution of mass loss from these areas. For a consistency check, the post-eruptive loops in the base-difference images on the left column of Figure \ref{Fig1} can be compared with the uniform color regions at the locations of the loops in the color maps (right column). This is because all pixels containing post-eruptive loops are assigned the same mass loss value, as stated above.

\subsection{Estimation of the Evacuated Mass} %%%%%%%%%%%%%%
  \label{S-Evacuated_Mass_method}

Under the isothermal hydrostatic equilibrium assumption, the density of the coronal plasma in the dimming region is radially stratified \citep{Aschwanden2004book}, 

\begin{equation} \label{Eq-density}
 N_e(z) = N_{e0} \, \exp \left[ -\frac{z/\lambda_p}{(z/R_\odot+1)} \right]
\end{equation}
The pressure scale height is given by,
\begin{equation}  \label{Eq-height_scale}
\lambda_{p} = \frac{2k_{B} T_e }{\mu  m_{H} g_\odot}
\end{equation}
where $T_e$ is the electron temperature, $k_{B}$ the Boltzmann constant, $\mu$ the mean molecular weight of the hydrogen ion, $m_{H}$ the hydrogen mass and $g_\odot$ the gravitational acceleration at the solar surface. We estimate the pressure scale height using the average temperature obtained from the DEM analysis, \ie~$T_e = \left<T\right>$.
From Equation (\ref{Eq-height_scale}), the average pressure scale height $\left<\lambda_{p}\right>$ obtained for the ROI pixels of the three events ranges between $\approx$\,75\,Mm to 90\,Mm.  

The basal electron density $N_{e0}$ can be obtained replacing Equation (\ref{Eq-density}) in Equation (\ref{EMdef})
\begin{equation} \label{Eq-EM-density}
 EM = \int_0^L N_e^2(z)\,dz = N_{e0}^2 \frac{\lambda_p}{2} \, I \left(\frac{\lambda_p}{2R\odot} \right) 
\end{equation}
\noindent where the length of the plasma column is set to $L=5\lambda_p$ in order to integrate through the bulk of the mass of the evacuated region, and we have defined the integral quantity,
\begin{equation} \label{Eq-Ialpha}
 I(\alpha) \equiv \int_0^{2L/\lambda_p}  \, {\rm exp} \left( - \frac{x}{\alpha x+1} \right) \,dx
\end{equation} 

The mass $M_i$ along the column of plasma corresponding to any given pixel $i$ can then be estimated as,
\begin{equation}  \label{Eq-lossmass}
 M_i = \mu\,m_H\,A_s\,\int_0^L N_{e,i}(z)\,dz =\mu\,m_H\,A_s\,\lambda_{p,i}\,N_{e0,i}\,I\left(\frac{\lambda_{p,i}}{R_\odot}\right)  
\end{equation}

\noindent where $A_s = \Omega_p\,d_s^2$ is the area of the patch in the Sun that is imaged by each pixel in the detector, with $\Omega_p$ being the solid angle subtended by the area of the pixel ($A_p$) with respect to the effective focal length ($f$) of the telescope (\ie~$\Omega_p=A_p/f^2$), and $d_s$ is the distance of the patch to the telescope. Note that $\Omega_p$ is a known intrinsic quantity of the telescope, while $d_s$ is known from the EUV image header. Finally, $N_{e0}$ can be obtained from Equation (\ref{Eq-EM-density}).

To determine the mass loss associated to a pixel $i$, the difference between the masses for the pre-eruption $M_{i}^{pre}$ and post-eruption $M_{i}^{pos}$ images is computed,

\begin{equation} \label{Eq-deltaM}
\Delta M_i = M_{i}^{pre} - M_{i}^{pos}
\end{equation}

The total evacuated mass, $\Delta M$, is computed as the sum of the mass loss from all the pixels of the dimming (fulfilling the condition given by Equation \ref{eq-condition}), plus the mass loss associated to the area covered by the post-eruptive loops. The contribution of this last term to the total evacuated mass represents 8\% for Event 1, 7\% for Event 2 and 11\% for Event 3. These values are obtained at the times of the images displayed in Figure \ref{Fig1}, when the post-eruptive loops are fully developed. This contribution is insensitive to the size of the red box defined to enclose the bright post-eruptive loops.

\begin{figure}    %%%%%%%%%%%%%%%%%% FIGURE 2
   \centerline{\hspace*{0.00\textwidth}
               \includegraphics[width=0.5\textwidth,clip=]{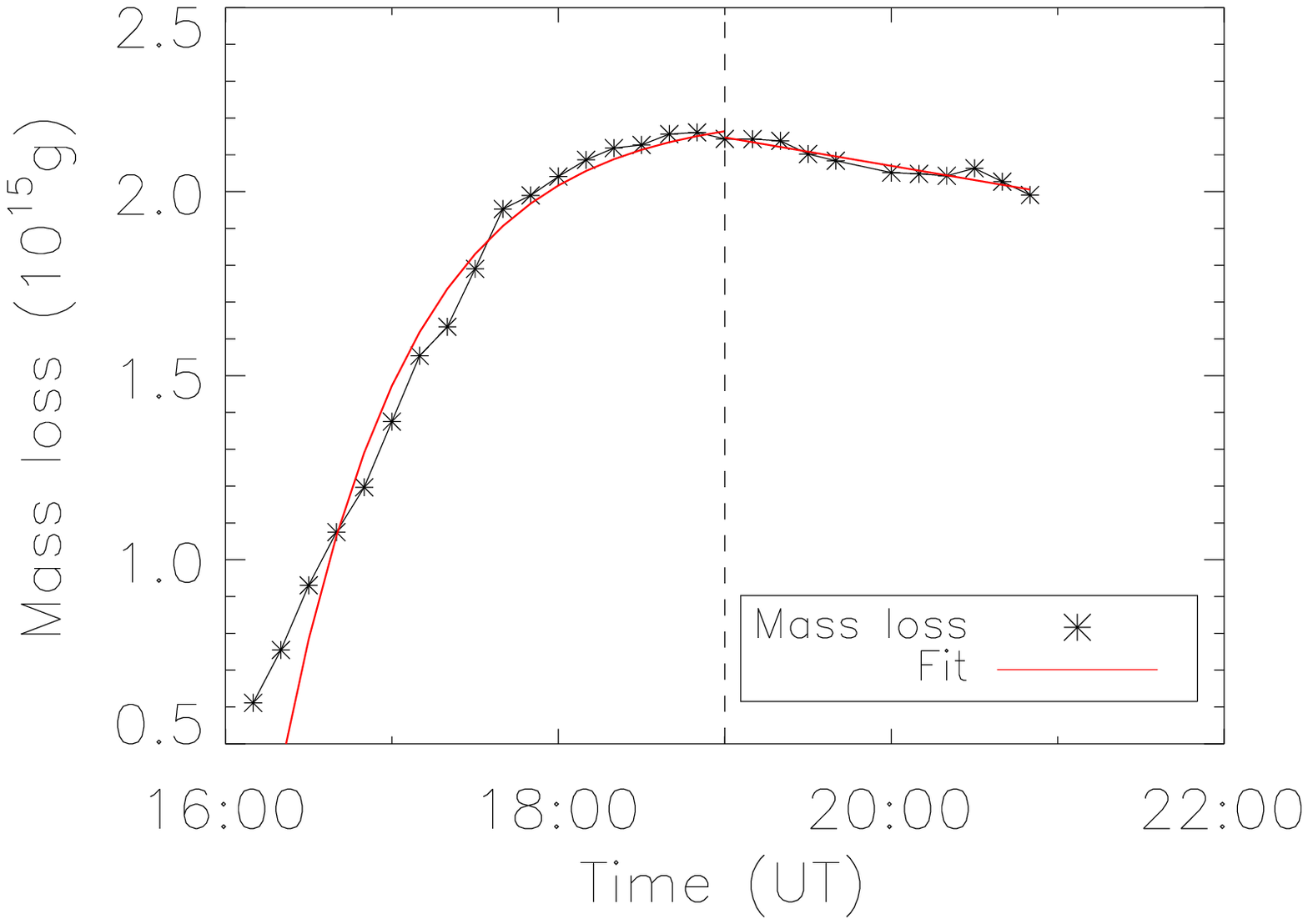}
               \hspace*{-0.01\textwidth}
               \includegraphics[width=0.5\textwidth,clip=]{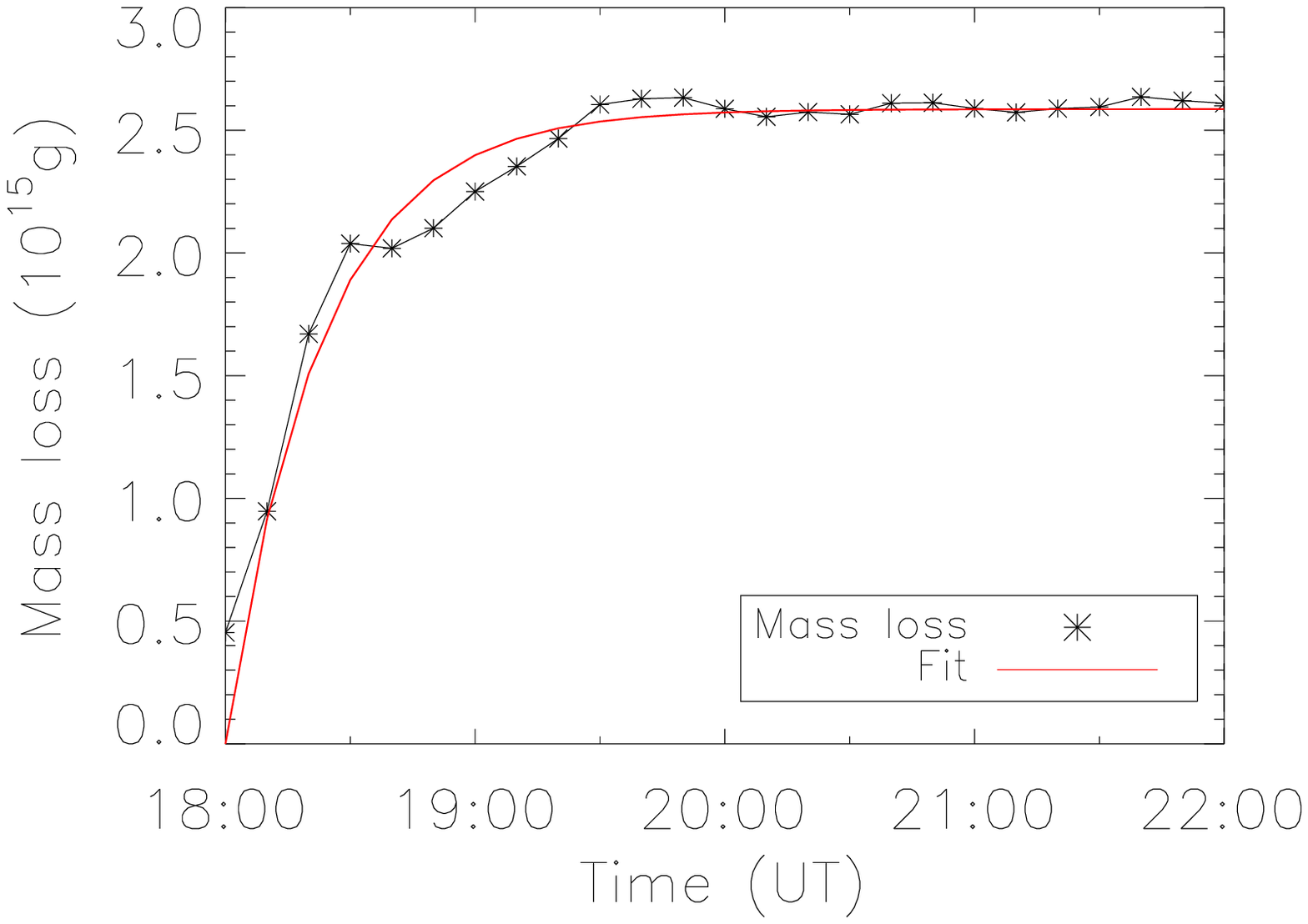}
               }
               \vspace{-0.34\textwidth}   % Shift close to the panel top 
     \centerline{\large \bf     % Includes the labels (here needs the color 
                                %   package, see beginning of this file)
      \hspace{0.09\textwidth}  \color{black}{(a)}
      \hspace{0.447\textwidth}  \color{black}{(b)}
         \hfill}
     \vspace{0.34\textwidth}    % Shift back to the panel bottom       
%------------------------------------------------------------------------------
 \centerline{\hspace*{0.00\textwidth}
     \includegraphics[width=0.5\textwidth,clip=]{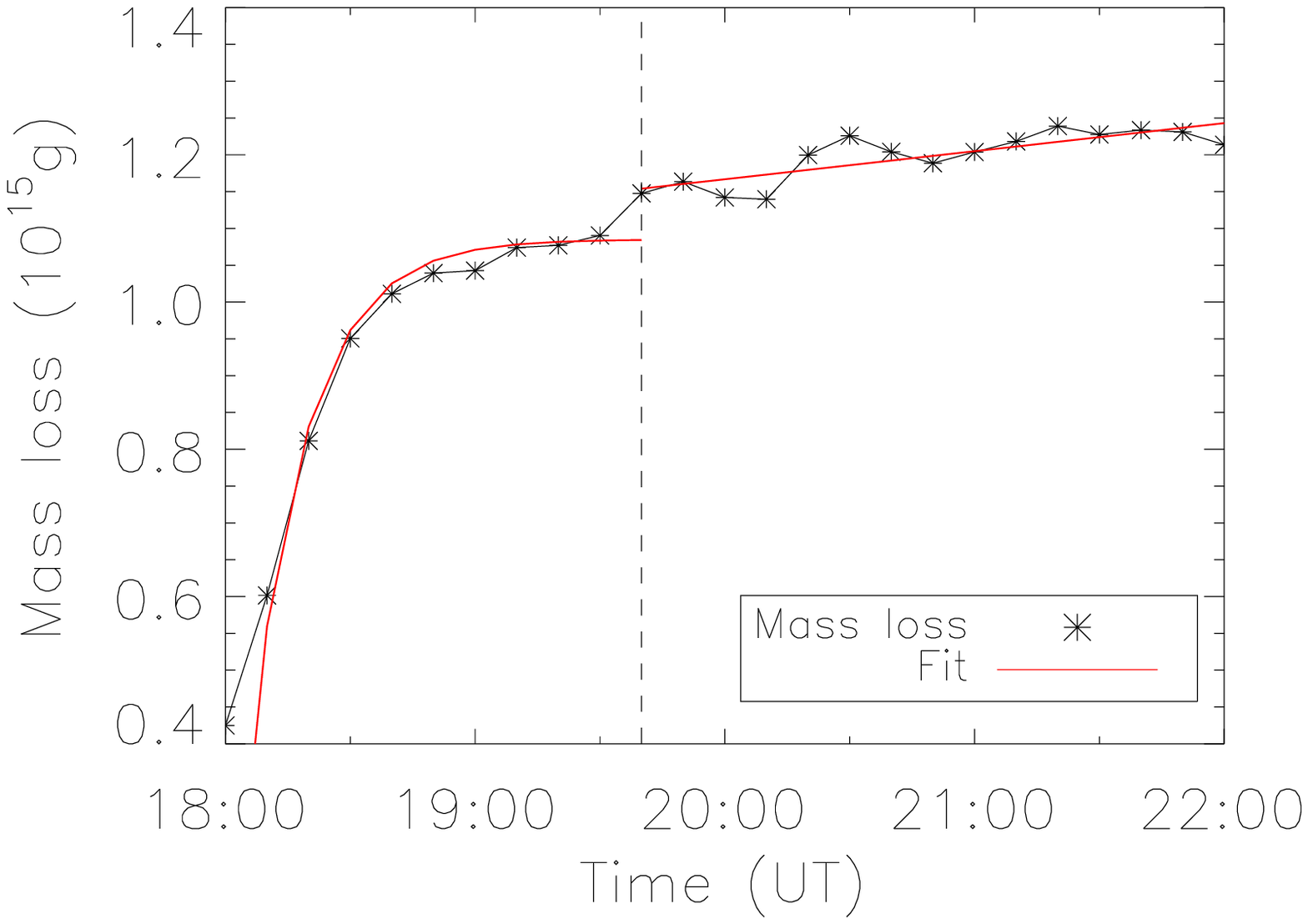}
     }
        \vspace{-0.34\textwidth}   % Shift close to the panel top 
     \centerline{\large \bf     % Includes the labels (here needs the color package)
      \hspace{0.345\textwidth} \color{black}{(c)}
         \hfill}
     \vspace{0.34\textwidth}    % Shift back to the panel bottom 
\caption{Temporal evolution of the mass loss from the low corona (asterisks joined with a line) for the events under study: 23 May 2010 (a), 7 August 2010 (b) and 30 November 2010 (c). Red solid lines represent the fit to the estimated mass values for each event, while the vertical dashed lines indicate the change of fitting function (see text). }
   \label{Fig2}
   \end{figure}

For each event, the evacuated mass $\Delta M$ was computed for a time series of images covering the evolution of the event. The resulting temporal behavior of the evacuated mass for the three analyzed events is shown in Figure \ref{Fig2}. The mass loss values determined every 10 minutes are represented by asterisks joined by lines. The red lines represent a fit to the values, comprised of two different functions. The first section represents the high evacuation rate in the initial stages, for which we use this Equation (\ref{Eq-fit_mass}):
\begin{equation} \label{Eq-fit_mass}
\Delta M(t) = A\,(1- e^{-Bt}) 
\end{equation}
This equation was proposed by \citet[][their Equation 11]{Colaninno2009} to fit the white-light mass evolution of eight CMEs.
We use a non-linear least squares fit, with initial guesses $[A,B]=[\Delta M_{top},\frac{3}{t_c}]$, where $\Delta M_{top}$ is the maximum value of the mass loss and $t_c$ is the time when the mass loss reaches 63\% of the $\Delta M_{top}$ value. 

The second function of the fit in Figure \ref{Fig2} is provided by a linear model, which represents the gradual recovery of the mass loss, as in Figure \ref{Fig2}a, or a slower evacuation phase, as in the case of Figure \ref{Fig2}c. For Event 2 in Figure \ref{Fig2}b, Equation (\ref{Eq-fit_mass}) was enough to describe the complete evolution during the analyzed time interval, because the mass loss approaches a constant value. The vertical dashed line in Figures \ref{Fig2}a and c indicates the change of fitting function. 

\section{CME Mass Determination} %%%%%%%%%%%%%%
  \label{S-CME_Mass}

It is known that the intensity registered by white-light coronagraphs is due to Thomson scattering of photospheric light by free electrons in the solar corona, with this intensity being the result of the individual contribution by all scattering elements along the LOS. This process is more efficient when the scattering electrons are located closer to the Thomson surface \citep[\eg][]{Vourlidas2006}. This surface is a sphere with a diameter equal to the Sun-observer distance. For COR2 white-light coronagraphs with a FOV $\leq$ 15 solar radii, the POS and the sphere can be considered as coincident. The emission decreases with the distance to this plane, producing an underestimation of the number of electrons of a CME when it is in fact propagating with a certain angle respect to the POS.

To determine the true direction of propagation of the CMEs, we apply the forward-modeling method developed by \citet{Thernisien2009} to simulate the CME simultaneously from three viewpoints to minimize errors in the fit. The three views are given by COR2-A, COR2-B and LASCO C2 images, whose background and structures not related to the CME are removed by subtracting a suitable pre-event image. Table \ref{T-mass-Comparison} shows some of the parameters obtained after applying this forward-modeling approximation to the three events. Column 1 indicates event number, while column 2 shows the Stonyhurst coordinates for the direction of propagation of the CMEs with respect to the Sun--Earth line. The angle subtended by the direction of propagation of the CMEs wih respect to the POS of COR2-A and COR2-B is given in columns 3 and 4 by $\omega_A$ and $\omega_B$ respectively. To measure the CME mass of the three events, we only used coronagraphic data of the spacecraft where the direction of propagation is closer to the respective POS, \ie~depending on whether the lowest value is given by $\omega_A$ or $\omega_B$. 

Once the direction of propagation of the CMEs is known, the total-brightness intensity images are converted to mass images by the Thomson scattering equations \citep{Billings1966}, following the method described in \citet{Vourlidas2010}. Here we are considering that all electrons lie on a plane at the corresponding angle $\omega$ relative to the POS, either $\omega_A$ or $\omega_B$.
The total mass of the CMEs is obtained by adding the contribution of the mass contained in each pixel that belongs to the CME. The boundary of the CME is then manually selected in the mass images, using the freehand ROI method. In this way, all the pixels enclosed in the ROI contribute to the CME mass. The CME mass values M$_{CME}$ presented in column 7 of Table \ref{T-mass-Comparison} are deduced from the last available image of the corresponding CME before it leaves the FOV of COR2.

To verify the validity of the results of our mass measurements from a single viewpoint, we apply the method presented in \citet{Colaninno2009}, which takes advantage of the two STEREO viewpoints to estimate the true CME masses. Using this approach, the true mass of a CME $M_T$ can be calculated using the equation $M_T =\left( M_A - M_B \right) / \left(f_m(\omega_A) - f_m(\omega_B) \right)$ where $M_A$ and $M_B$ are the masses determined from COR2-A and COR2-B respectively, considering the POS assumption ($\omega = 0$). The angles $\omega_A$ and $\omega_B$ represent the angular separation between the CME direction of propagation and the respective POS of COR2-A and COR2-B. The function $f_m(\omega)$ is the ratio of the brightness of an electron at an angle $\omega$ relative to its brightness on the POS: $f_m(\omega) = B_e(\omega) / B_e(\omega = 0)$. 

The resulting masses obtained from the application of the method by \cite{Colaninno2009} are in good agreement with those found from single-spacecraft data. Differences in the masses obtained by the two methods are in the order of $\sim$6\% for events 1 and 3. However, for Event 2, whose CME is a partial halo from COR2-B, the discrepancy is considerably high. A similar result is also reported by these authors, who did not find a good agreement for the case of a partial halo CME event, indicating a limitation in the accuracy of their method when the CME travels far from the POS of at least one of the spacecraft. Because of this shortcoming for Event 2 and the small differences for Events 1 and 3, we prefer to list in Table \ref{T-mass-Comparison} only the CME mass deduced from single-spacecraft data.

\section{Comparison of CME Mass and Mass Loss in the Low Corona} %%%%%%%%%%%%%%
  \label{S-Mass-Comparison}

In this section, we present the results that arise from the comparison between (i) the masses determined for the CMEs from white-light coronagraphic observations, and (ii) the evacuated mass from the associated dimming regions from EUV images. Column 8 of Table \ref{T-mass-Comparison} displays the obtained values of the mass loss in the three dimming regions (M$_{EUV}$), while column 9 shows the percentage ratio of  M$_{EUV}$/M$_{CME}$ (Ratio).  

\begin{figure}    %%%%%%%%%%%%%%%%%% FIGURE 3
                                % includes the three top panels 
   \centerline{\hspace*{0.00\textwidth}
               \includegraphics[width=0.5\textwidth,clip=]{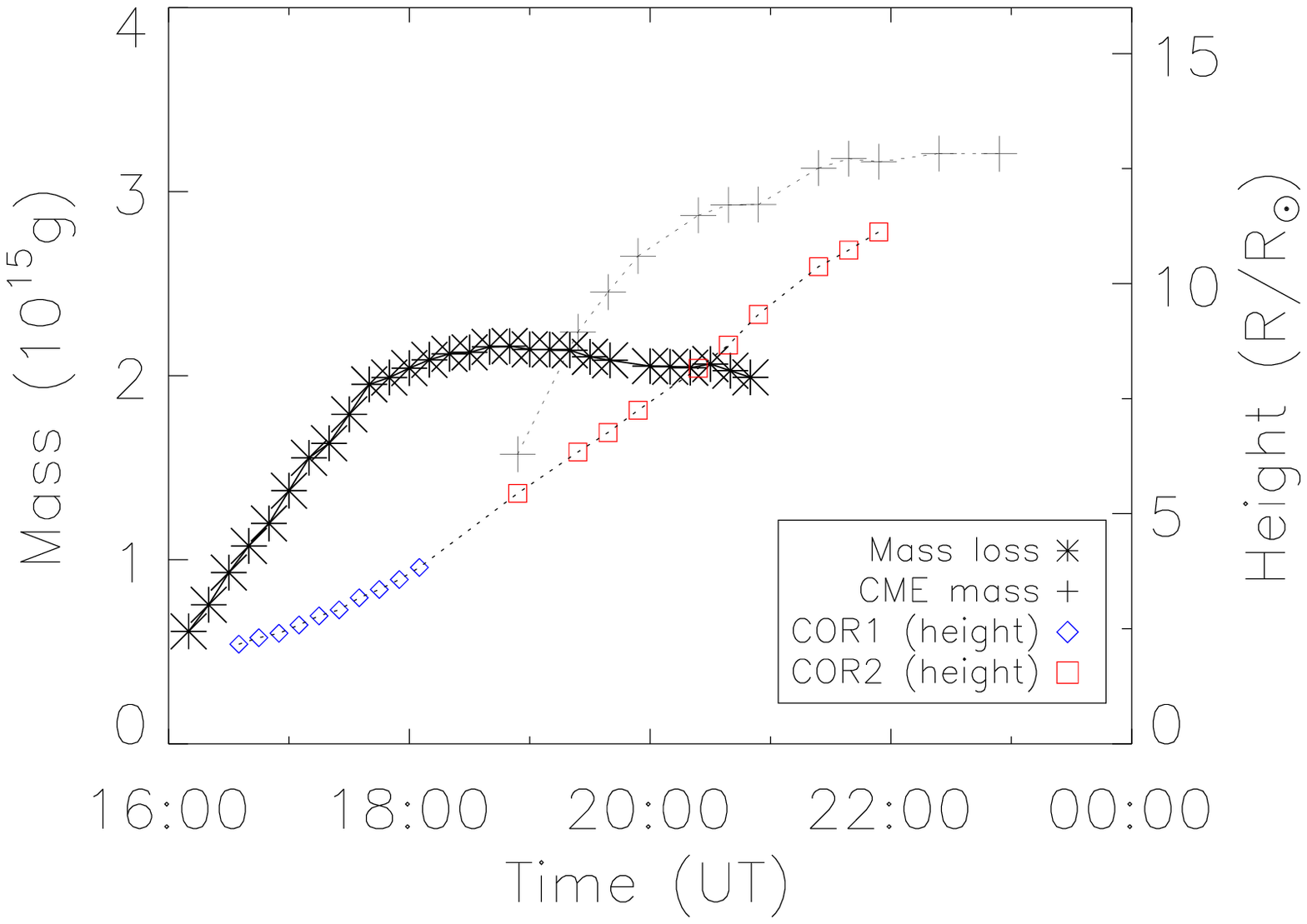}
               \hspace*{-0.01\textwidth}
               \includegraphics[width=0.5\textwidth,clip=]{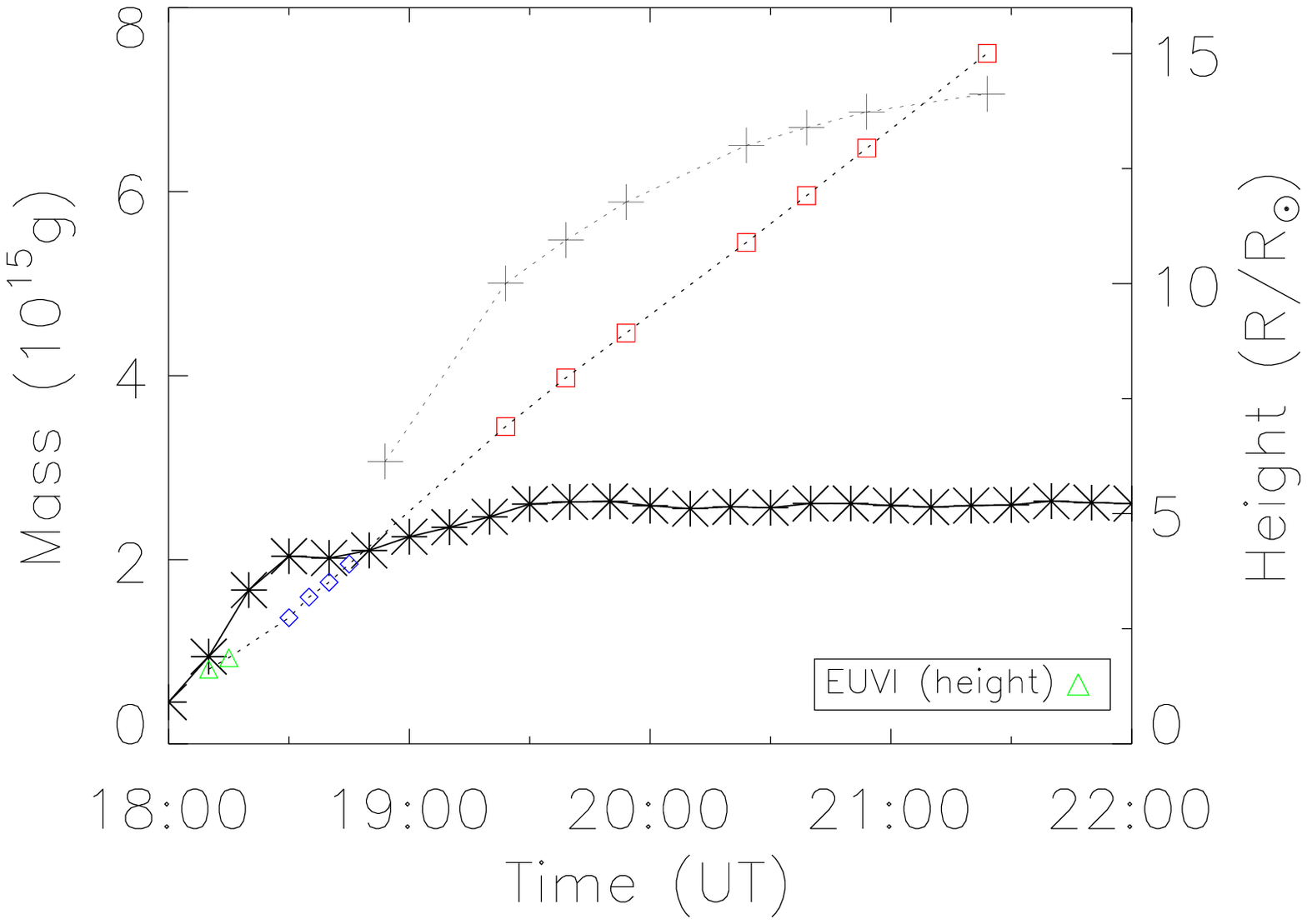}
               }
               \vspace{-0.34\textwidth}   % Shift close to the panel top 
     \centerline{\large \bf     % Includes the labels (here needs the color 
                                %   package, see beginning of this file)
      \hspace{0.075\textwidth}  \color{black}{(a)}
      \hspace{0.433\textwidth}  \color{black}{(b)}
         \hfill}
     \vspace{0.34\textwidth}    % Shift back to the panel bottom       
%------------------------------------------------------------------------------
 \centerline{\hspace*{0.00\textwidth}
     \includegraphics[width=0.5\textwidth,clip=]{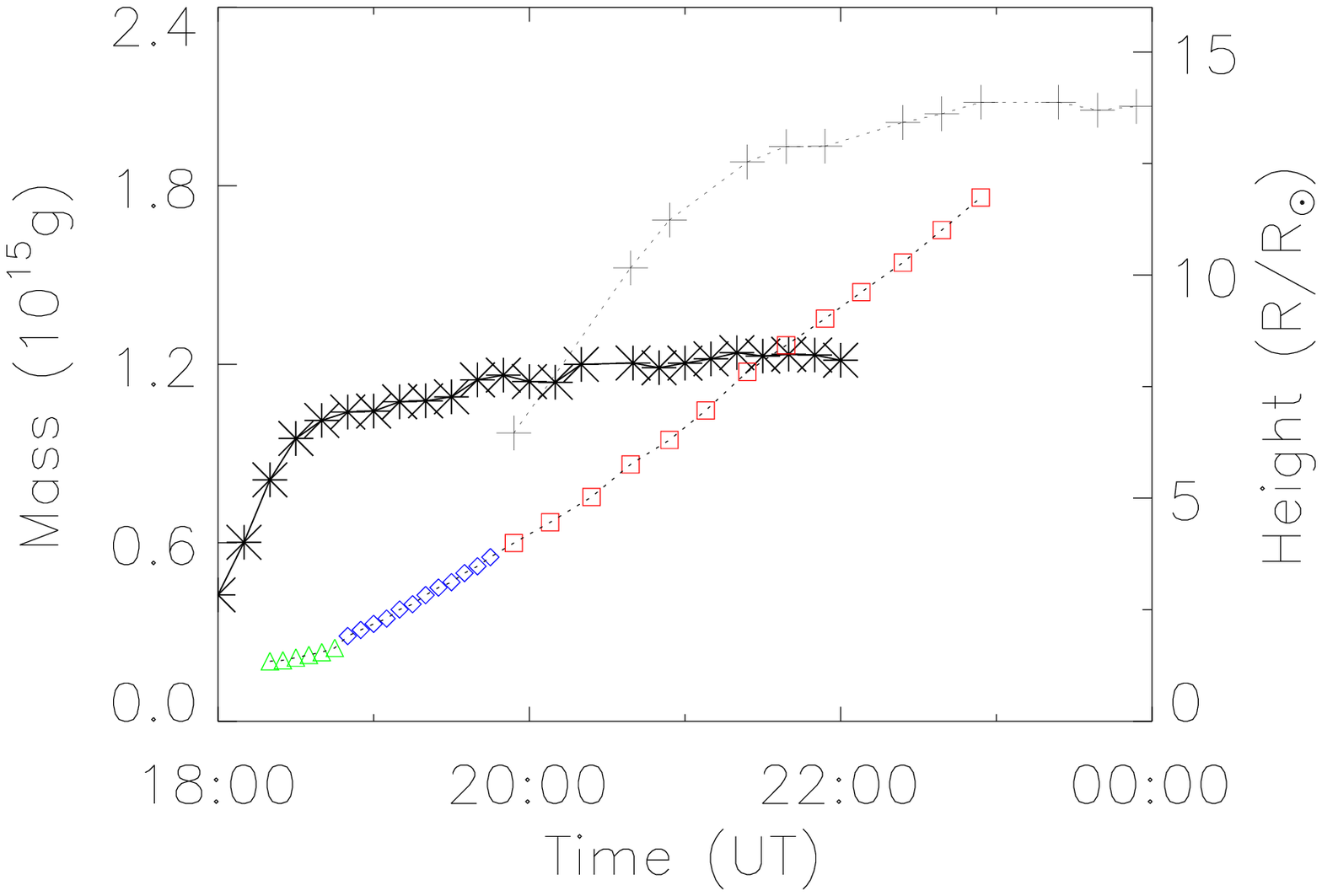}
     }
        \vspace{-0.34\textwidth}   % Shift close to the panel top 
     \centerline{\large \bf     % Includes the labels (here needs the color package)
      \hspace{0.33     \textwidth} \color{black}{(c)}
         \hfill}
     \vspace{0.37\textwidth}    % Shift back to the panel bottom 
\caption{Temporal evolution of EUV mass loss (asterisks) and white-light CME mass (plus signs) for the events under study: 23 May 2010 (a), 7 August 2010 (b) and 30 November 2010 (c). The deprojected height-time points corresponding to the CME's leading edge are also shown, with green triangles representing the EUVI measurements, blue diamonds those of COR1, and red squares those of COR2.}
\label{Fig3}
\end{figure}

Figure \ref{Fig3} shows the temporal evolution of the EUV mass loss in comparison with that of the associated CME mass for the three events. Both curves present an initial fast rise phase followed by a considerably slower variation in time. The evident temporal offset between the mass loss and the CME mass temporal profiles can be attributed to the mass of the CME emerging into the COR2's FOV as the CME propagates outward from behind the occulter.

In addition, Figure \ref{Fig3} displays the deprojected height (\ie~corrected by the angle of propagation $\omega$, addressed in the previous section) of the CMEs’ leading edge as a function of time, determined from EUVI, COR1 and COR2 observations. For Event 1, it was not possible to clearly identify the CME front in the low corona. However, a second-order profile is evidenced from the COR1 points. Event 2 exhibits an almost linear profile, even at low coronal heights. The kinematic evolution of Event 3 shows an accelerated profile for the first set of points, while it can be considered linear after $\approx$\,3 R$_\odot$. For completeness, the acceleration and the speed at the height of the last measurement, obtained from a second-order fit are shown in columns 5 and 6 of Table \ref{T-mass-Comparison}. From Figure \ref{Fig3}, we can also infer that the highest rate of mass evacuation from the dimming regions takes place when the CMEs are traveling through the FOV of EUVI and COR1. The evacuation of plasma continues even when de CME's leading edge is at COR2 coronal heights, persisting up to $\approx$\,5\,R$_\odot$ for Event 1 and $\approx$\,7\,R$_\odot$ for Event 2. Event 3 presents mass loss when the CME is beyond 10\,R$_\odot$, although at a considerable lower pace. However, it must be noted that the mass supplied from the low corona during the propagation of the three CMEs in the COR2’s FOV is small, in comparison with the difference between M$_{CME}$ and M$_{EUV}$. This difference may be in part due to the mass that the CMEs may add during their propagation in the higher lying corona they pass through, as summarized in Section \ref{S-Summary}.

\begin{table}
\caption{Various parameters obtained for events 1\,--\,3. Column 1 refers to the event identification number, and Columns 2\,--\,4 correspond to parameters obtained by the forward-modeling approximation. Columns 5 and 6 show the CME acceleration and speed at the height of the last measurement, obtained from a second-order fit. Columns 7 and 8 are the masses deduced from the white-light and EUV data respectively, while column 9 shows the consequent ratio between them.}
\label{T-mass-Comparison}
\begin{tabular}{lcccccccl r@{.}l c}
\hline
 & Prop. dir.& $\omega_A$ & $\omega_B$ & Acc. & Speed & M$_{CME}$ & M$_{EUV}$ & Ratio \\
\# & [lat., long.] & (deg.) & (deg.) & (m s$^{-2}$)& (km s$^{-1}$) & ($10^{15}$g) & ($10^{15}$g) & (\%) \\
(1) & (2) & (3) & (4) & (5) & (6) & (7) & (8) & (9) \\
\hline \\
1 & [N08,015] & 34 & 5 & 10.8 & 445.68  & 3.21  & 2.16  & 67.2                                  \\
2 & [N05,317] & 32 & 62 & -8.2 & 761.97 & 7.06 & 2.63  & 37.2                                   \\
3 & [N03,328] & 27 & 39 & 17.4 & 600.80 & 2.11 & 1.23  & 58.3                                   \\
\hline
\end{tabular}
\end{table}

% Acá va la figura 4
\begin{figure}    %%%%%%%%%%%%%%%%%% FIGURE 4
    \centerline{\hspace*{0.00\textwidth}
               \includegraphics[width=0.5\textwidth,clip=]{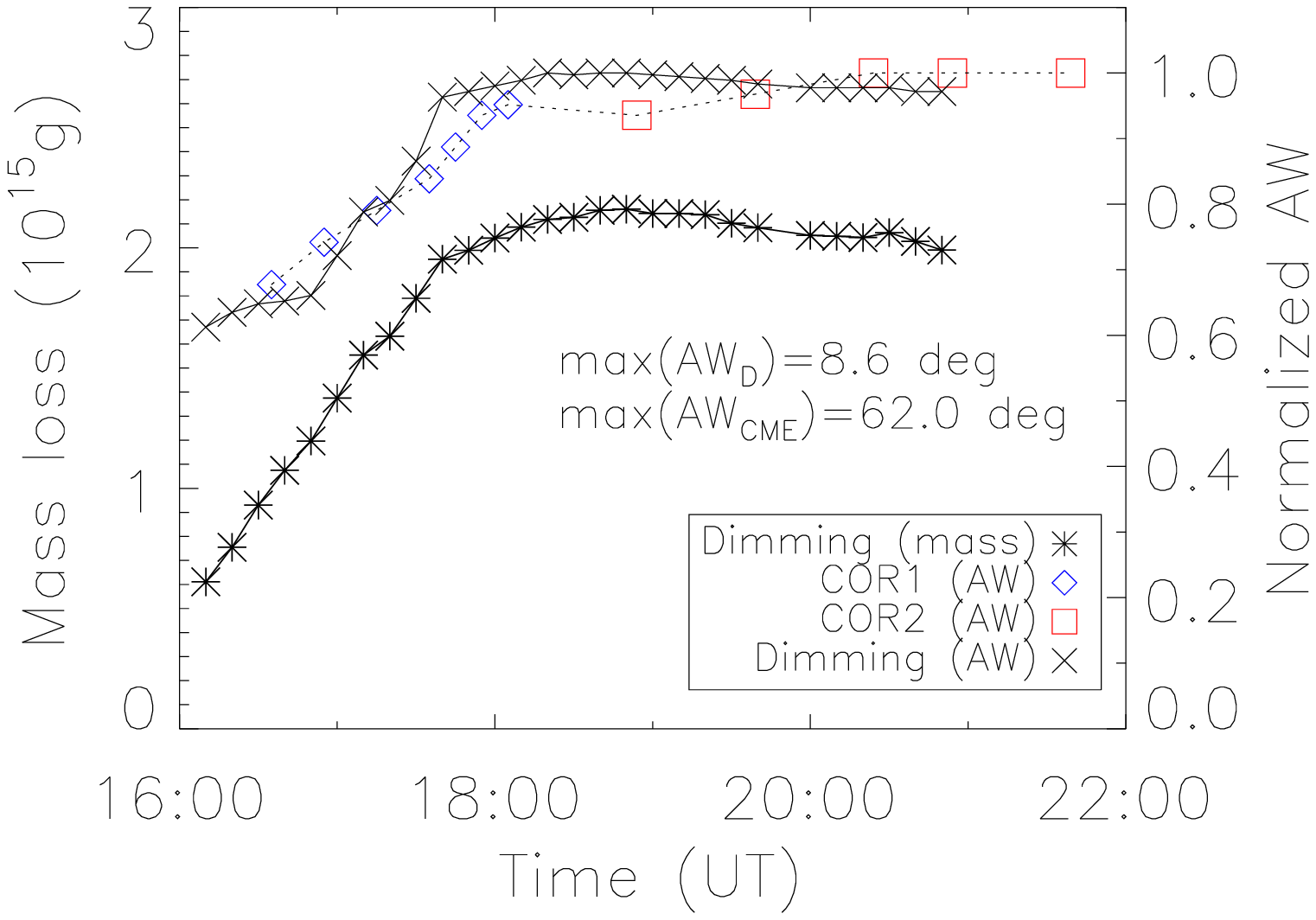}
               \hspace*{0.00\textwidth}
               \includegraphics[width=0.5\textwidth,clip=]{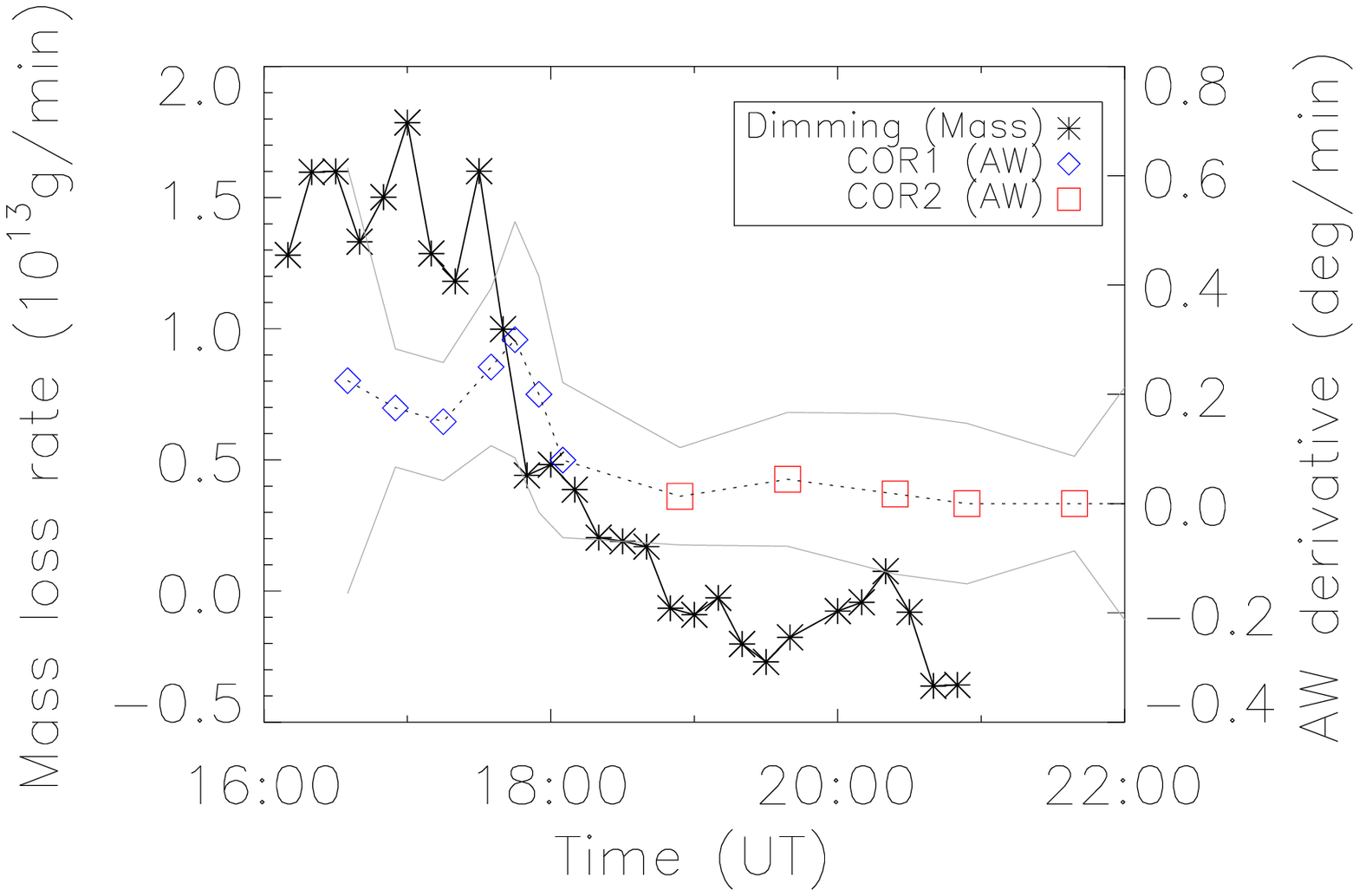}
               }
               \vspace{-0.36\textwidth}   % Shift close to the panel top 
     \centerline{\large \bf     % Includes the labels (here needs the color 
                                %   package, see beginning of this file)
      \hspace{0.06 \textwidth}  \color{black}{(a)}
      \hspace{0.45\textwidth}  \color{black}{(b)}
         \hfill}
     \vspace{0.36\textwidth}    % Shift back to the panel bottom       
  
%------------------
   \centerline{\hspace*{0.00\textwidth}
               \includegraphics[width=0.5\textwidth,clip=]{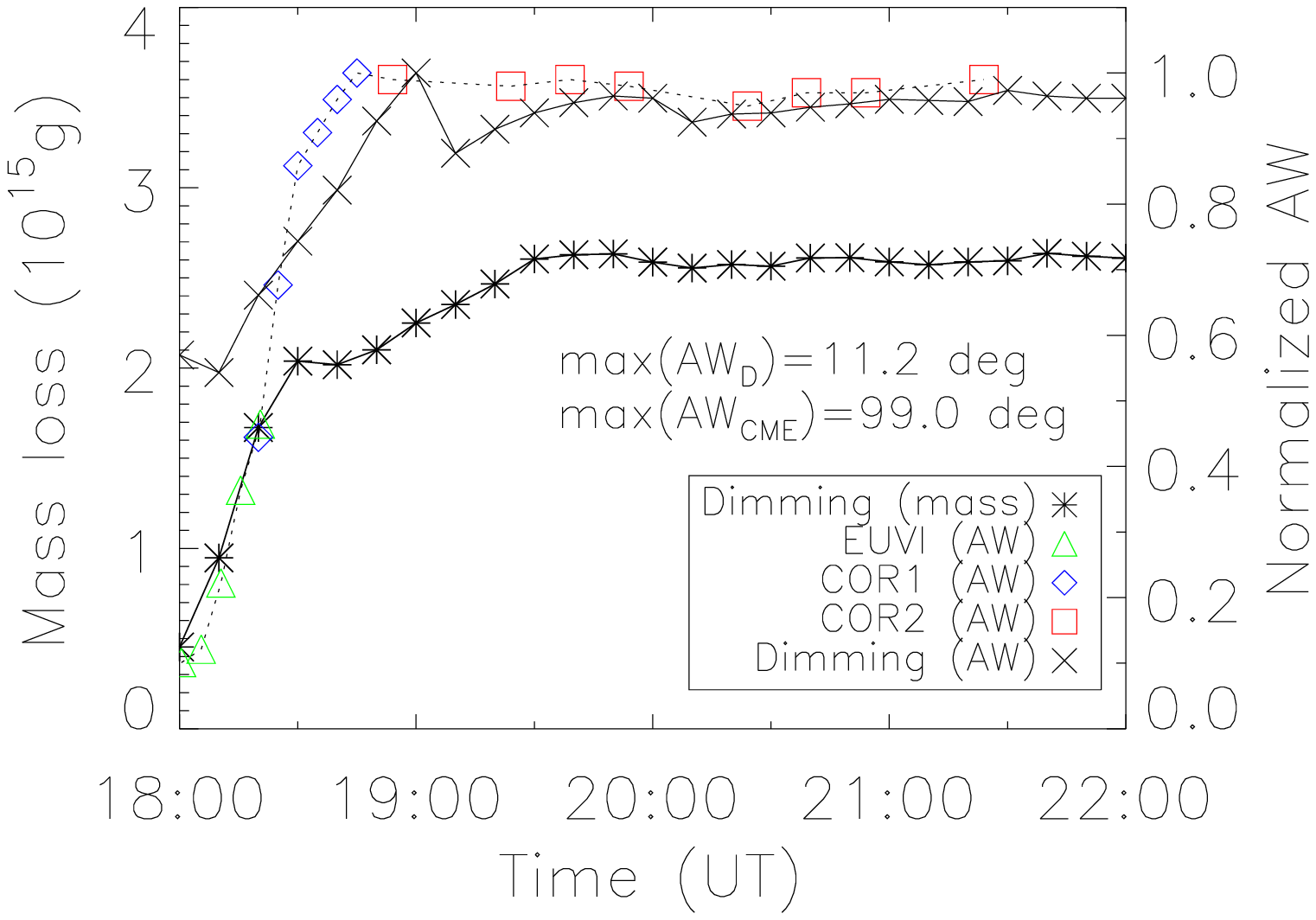}
               \hspace*{0.00\textwidth}
               \includegraphics[width=0.5\textwidth,clip=]{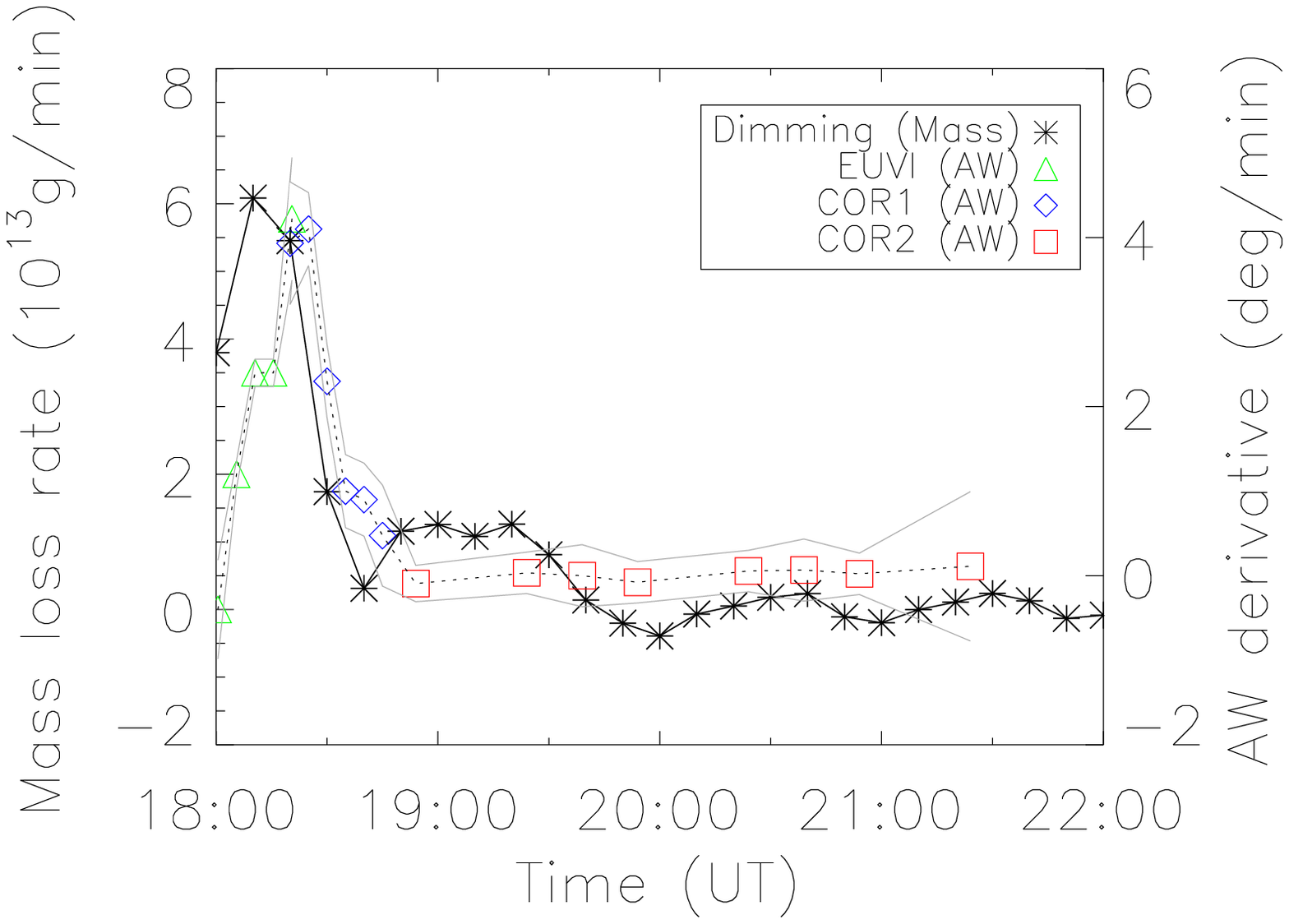}
              }
     \vspace{-0.36\textwidth}   % Shift close to the panel top 
     \centerline{\large \bf     % Includes the labels (here needs the color package)
      \hspace{0.06\textwidth} \color{black}{(c)}
      \hspace{0.45\textwidth}  \color{black}{(d)}
         \hfill}
     \vspace{0.36\textwidth}    % Shift back to the panel bottom 
%------------------------     
 \centerline{\hspace*{0.00\textwidth}
               \includegraphics[width=0.5\textwidth,clip=]{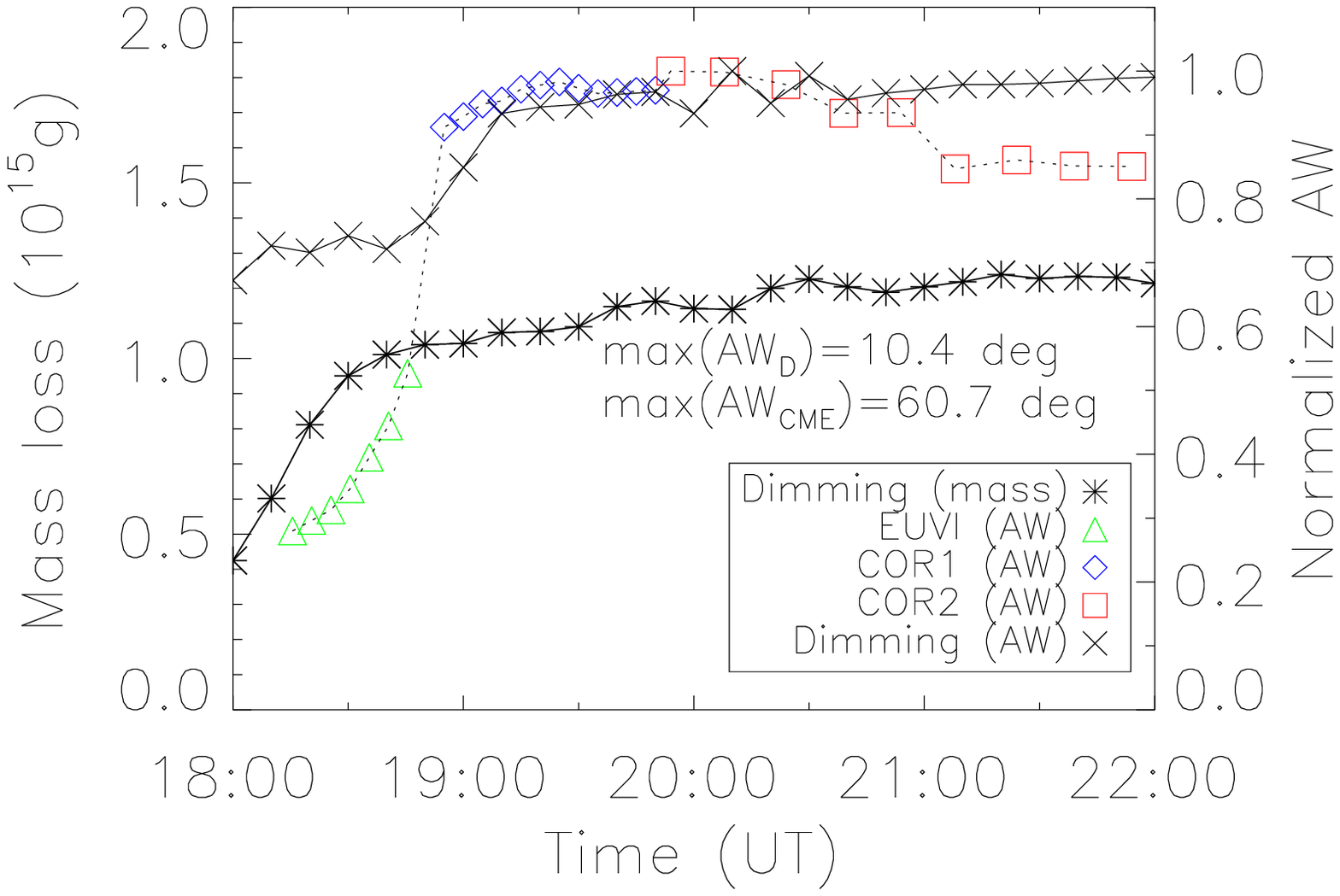}
               \hspace*{0.00\textwidth}
               \includegraphics[width=0.5\textwidth,clip=]{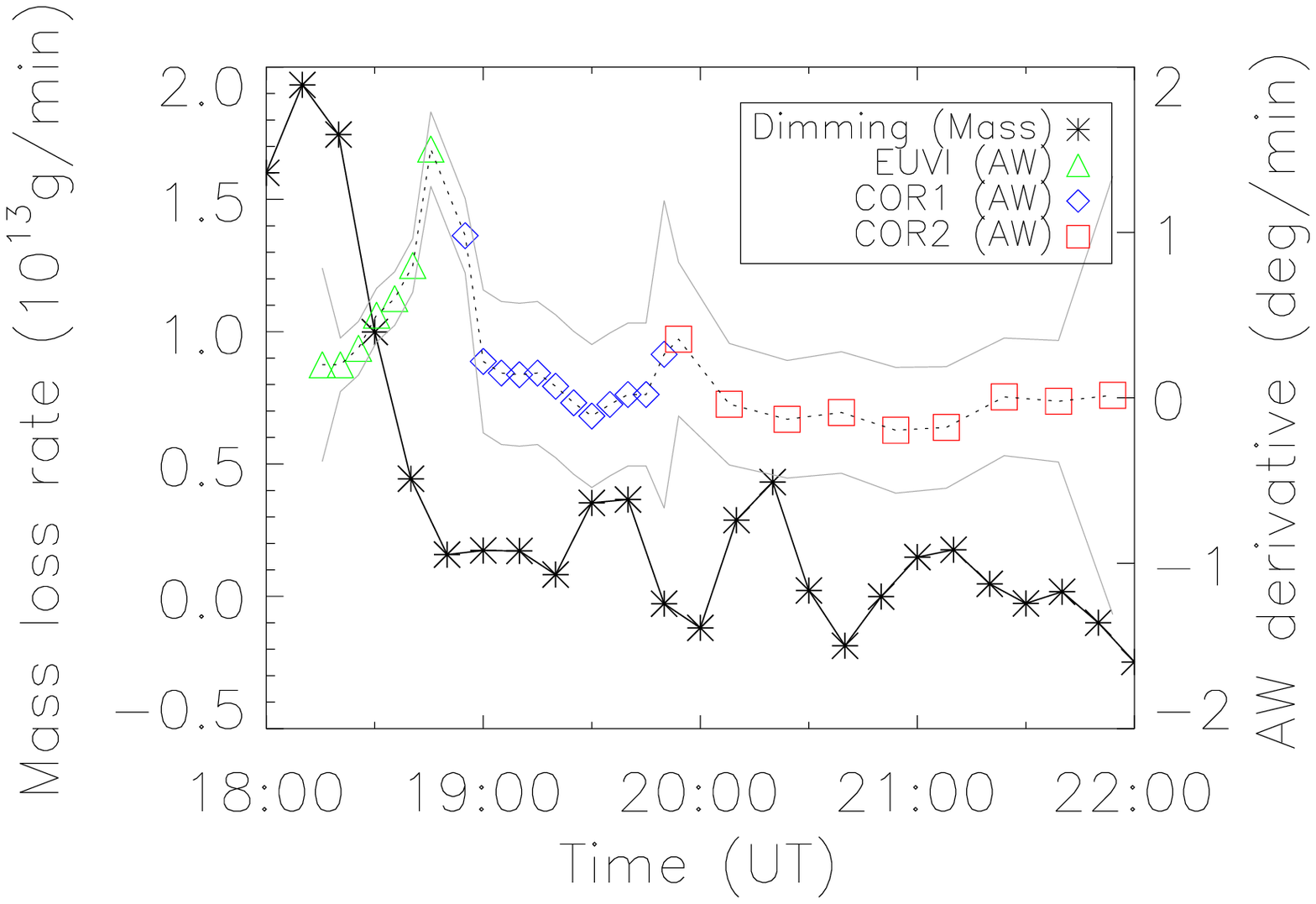}
               }
               \vspace{-0.36\textwidth}   % Shift close to the panel top 
     \centerline{\large \bf     % Includes the labels (here needs the color 
                                %   package, see beginning of this file)
      \hspace{0.06 \textwidth}  \color{black}{(e)}
      \hspace{0.45\textwidth}  \color{black}{(f)}
         \hfill}
     \vspace{0.36\textwidth}    % Shift back to the panel bottom  

\caption{Left column: temporal evolution of the normalized AW of the dimming area defined by the DEM (crosses), of the CME in EUVI (green triangles), COR1 (blue diamonds) and COR2 (red squares), for Event 1 (a), Event 2 (b) and Event 3 (c). The maximum AW values used to normalize the AWs of the dimmings and of the CMEs are shown in the plots as max($AW_{D}$) and max($AW_{CME}$) respectively. The measurements of the AW of the CME in the three instruments are joined by a dotted line. For comparison, the low coronal evacuated mass is also displayed (asterisks). Right column: derivative of the CME AW at the times of the AW measurements using the same symbols as in (a), (b) and (c). The band delimited by the gray continuous lines in figures (b), (d), and (f) shows the uncertainty in the AW derivatives. For comparison, the derivative of the evacuated mass is also displayed (asterisks).}

\label{Fig4}
\end{figure}

To investigate the temporal evolution of the CMEs' angular width (AW) from the low corona and up to the reach of COR2's FOV, we have determined the angular span of the CME in EUVI, COR1 and COR2 images. In the left column of Figure \ref{Fig4} we show the results of these measurements determined from EUV and white-light data in comparison with the AW of the dimming regions and the mass loss from the low corona. All AWs have been normalized to the respective maximum values, displayed in the plots as max($AW_{D}$) and max($AW_{CME}$). 

The determination of the CMEs' AW is carried out in images provided by STEREO A or B depending on which spacecraft detects the CME traveling closer to its POS. 

We calculate the uncertainties in the CME AW estimation as $\sigma_{AW} = 2\,(\sigma_{m} / R_{oc})$ (in radians), where ($\sigma_{m}$) is the error in the measurement of the position of a CME's flank, and R$_{m}$ the height of the measurement. The considered values of ($\sigma_{m}$) are 0.01\,R$_\odot$ for EUVI, 0.04\,R$_\odot$ for COR1, and 0.17\,R$_\odot$ for COR2; at a height R$_{m}$ of 1\,R$_\odot$ for EUVI and at the border of the occulter for each coronagraph. These values of R$_m$ correspond to the heights at which the measurements of the CME's flanks are approximately performed. The errors in the AW of the CMEs are thus estimated to be 1.1$^{\circ}$ for EUVI, 3.0$^{\circ}$ for COR1 and 7.8$^{\circ}$ for COR2.

The AW of the dimming, on the other hand, is determined by considering the dimming as a disk of area $A_D$ and radius $R_D=\sqrt{A_D /\pi}$. The area on the spherical surface of the Sun, covered by a pixel $i$ is determined as $a_{d,i} = A_s /(cos(\rho_i))$, with $\rho$ being the angle between the normal to the local surface and the direction to the observer, and $A_s$ the same as defined in Section \ref{S-Evacuated_Mass_method}. 
The area of the dimming ($A_D$) is obtained after summing the area on the surface of the Sun contained by all pixels that belong to the dimming ($A_D= \sum a_{d}$). The AW of the dimming region ($AW_D$) is then given by the relation $AW_D=2\arctan(R_D/1\,R_\odot)$. 

The profiles of normalized AW of the dimming regions and of the CMEs in the left column of Figure \ref{Fig4} exhibit a good concordance. In turn, both AW profiles are also in agreement with the temporal evolution of the mass loss. The AW for both the CME and the dimming increases during the fastest phase of mass evacuation. Likewise, both AWs become stable roughly at the time the mass evacuation ceases. Event 2, which is associated with an M-class flare, presents the maximum change of normalized AW at COR1 heights, from $\approx$\,0.4 to 1.0, while in the FOV of EUVI it changes from $\approx$\,0.1 to 0.4. Event 1, related to a B-class flare, shows a change of $\approx$\,0.7\,--\,1.0 in COR1. As for Event 3, which could not be associated to any flare, its change in normalized AW is merely from $\approx$\,0.9 to 1.0 in COR1, while most of its expansion takes place in EUVI, from $\approx$\,0.3 to at least 0.5. The analysis of these three events suggests a possible relationship between the intensity of the associated flare and the relative expansion of the CME. On the one hand, the more intense the flare, the larger the change of normalized AW at COR1 heights is. On the other, the less energetic the flare (actually Event 3 lacks a flare), the lower the height at which most of the CME expansion takes place. Naturally, this result cannot be generalized and may apply only to this small set of analyzed events.

The right column in Figure \ref{Fig4} shows the derivative of the CMEs' AW together with that of the evacuated mass, as a function of time. The AW and mass loss derivatives are obtained by applying a 3-point Lagrangian interpolation. There is a good overall agreement in the general behavior of the derivatives of AW and mass loss profiles, both in time and relative amplitude. The largest AW rates match in time with the growth of mass loss rate for events 1 and 2, while both rates fall down and become stable simultaneously. For Event 3 there is an initial disagreement in time between both, AW and mass loss derivatives. In fact, the highest mass loss rates occur when the opening of the EUV loops starts to develop, while the peak of the AW derivative takes place when the mass loss rate has significantly decreased. After that, both profiles fall down slowly, with the AW derivative profile becoming more stable. In correspondence with the interpretation of the plots in the first column of the figure, the highest AW rates are achieved by Event 2. Events 1 and 3 reach lower peak values of AW rate, with the expansion of Event 1 being more sustained over time, most likely because we were not able to perform AW measurements in EUVI for this event.

\section{Temporal Evolution of the Evacuated Mass and X-ray Flux}
\label{S-Xray}

\begin{figure}    %%%%%%%%%%%%%%%%%% FIGURE 5
    \centerline{\hspace*{0.00\textwidth}
               \includegraphics[width=0.5\textwidth,clip=]{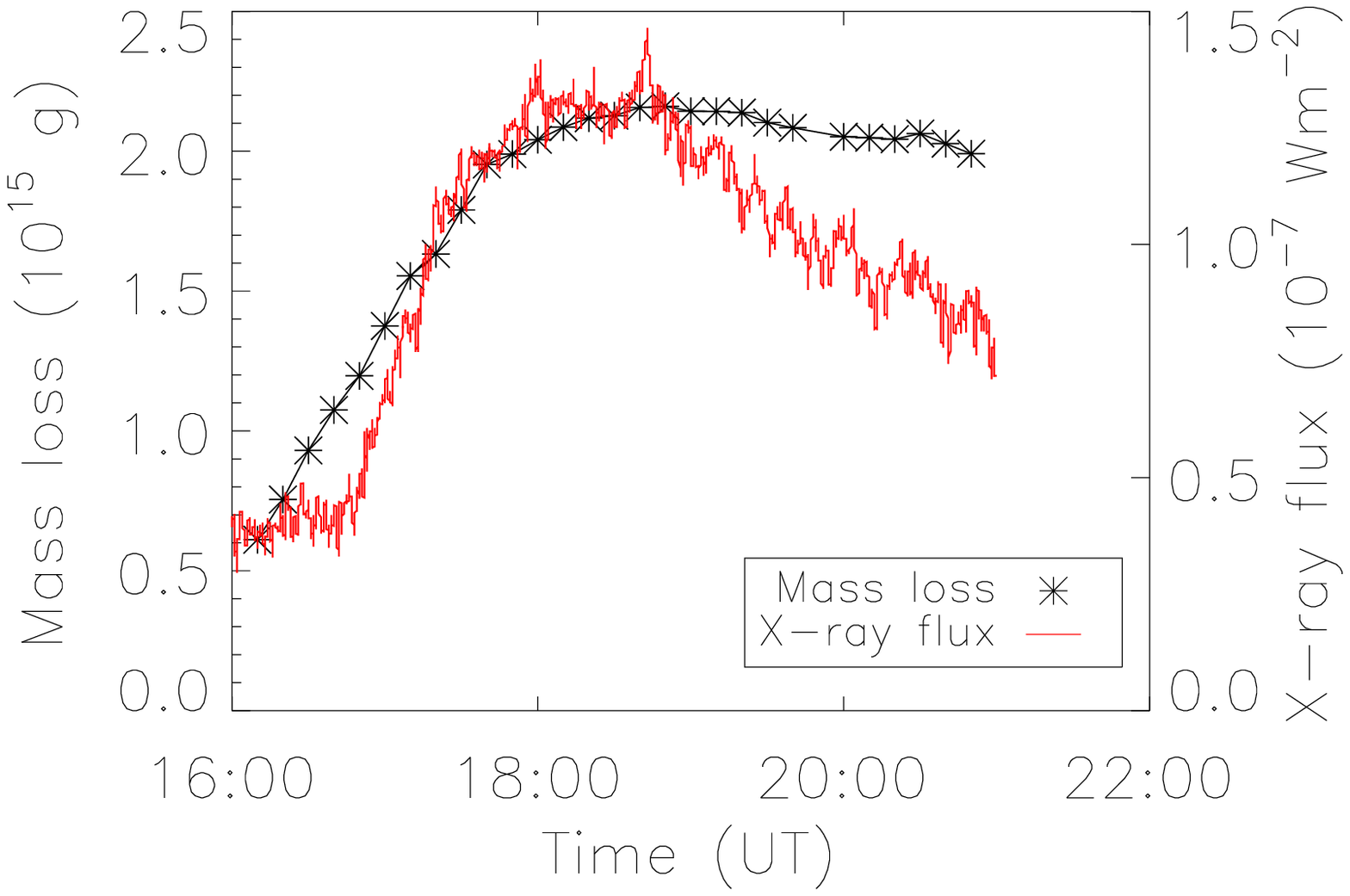}
               \hspace*{0.00\textwidth}
               \includegraphics[width=0.5\textwidth,clip=]{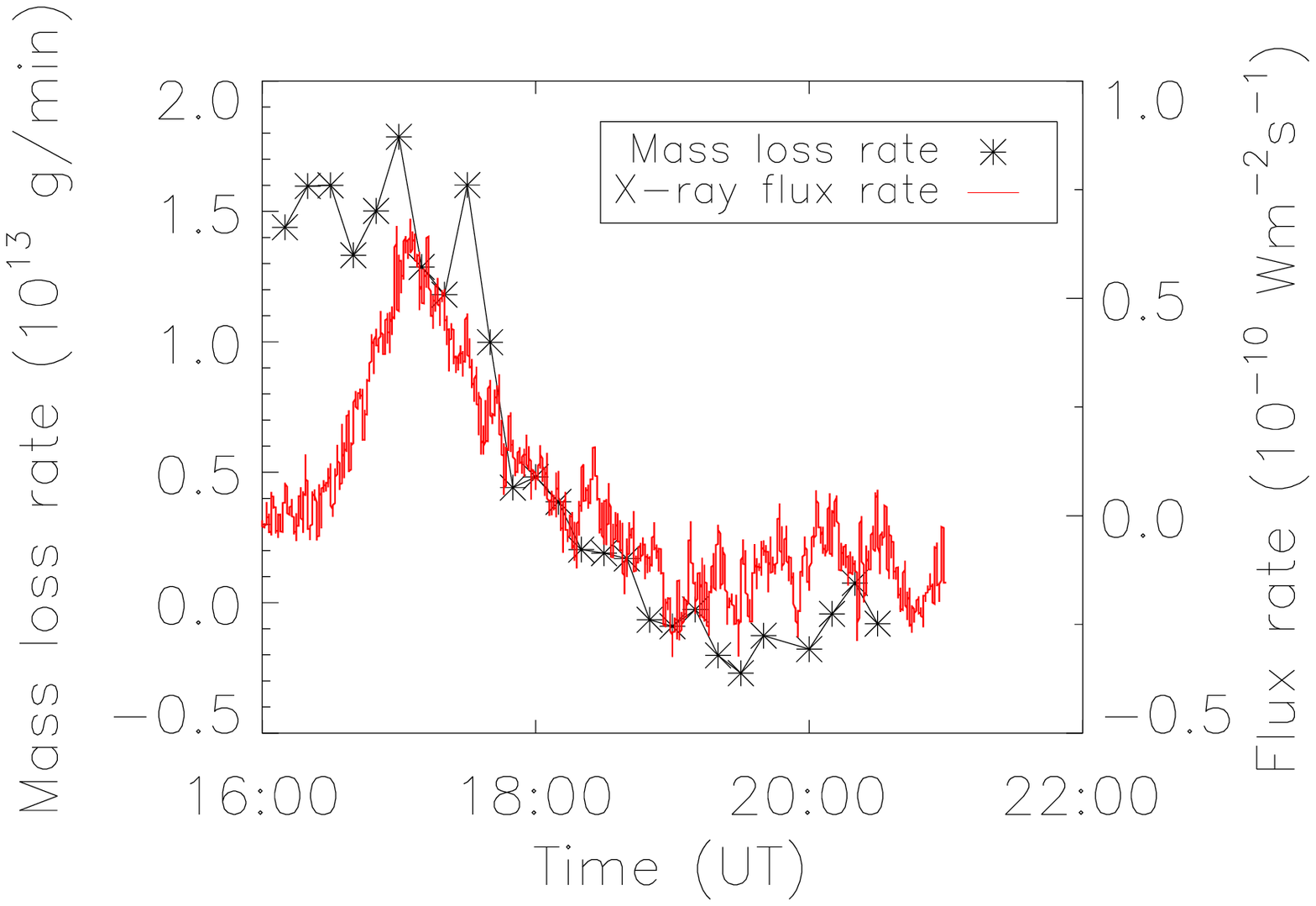}
               }
               \vspace{-0.36\textwidth}   % Shift close to the panel top 
     \centerline{\large \bf     % Includes the labels (here needs the color 
                                %   package, see beginning of this file)
      \hspace{0.06 \textwidth}  \color{black}{(a)}
      \hspace{0.45\textwidth}  \color{black}{(b)}
         \hfill}
     \vspace{0.36\textwidth}    % Shift back to the panel bottom       
  
%------------------
   \centerline{\hspace*{0.00\textwidth}
               \includegraphics[width=0.5\textwidth,clip=]{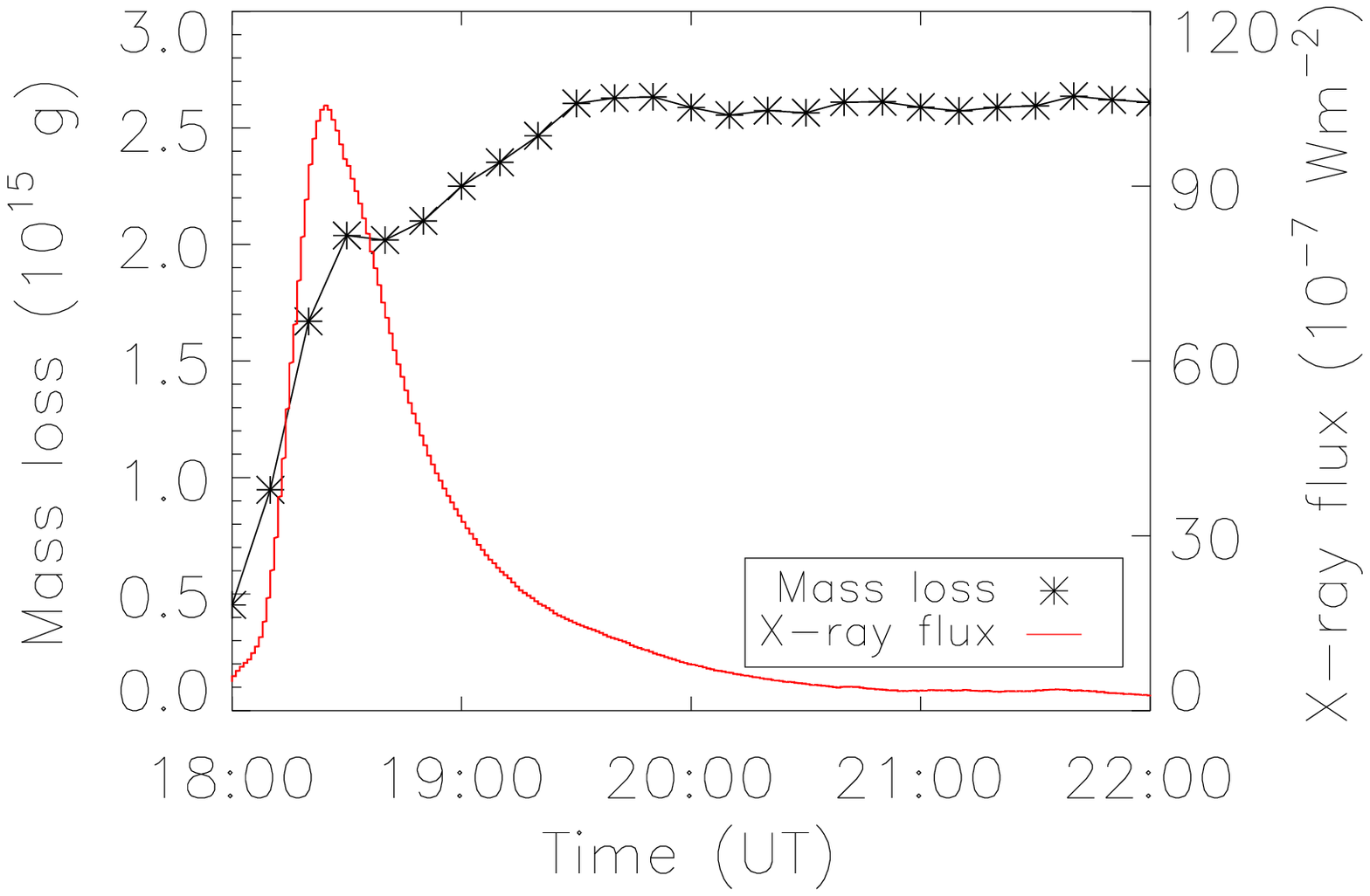}
               \hspace*{0.00\textwidth}
               \includegraphics[width=0.5\textwidth,clip=]{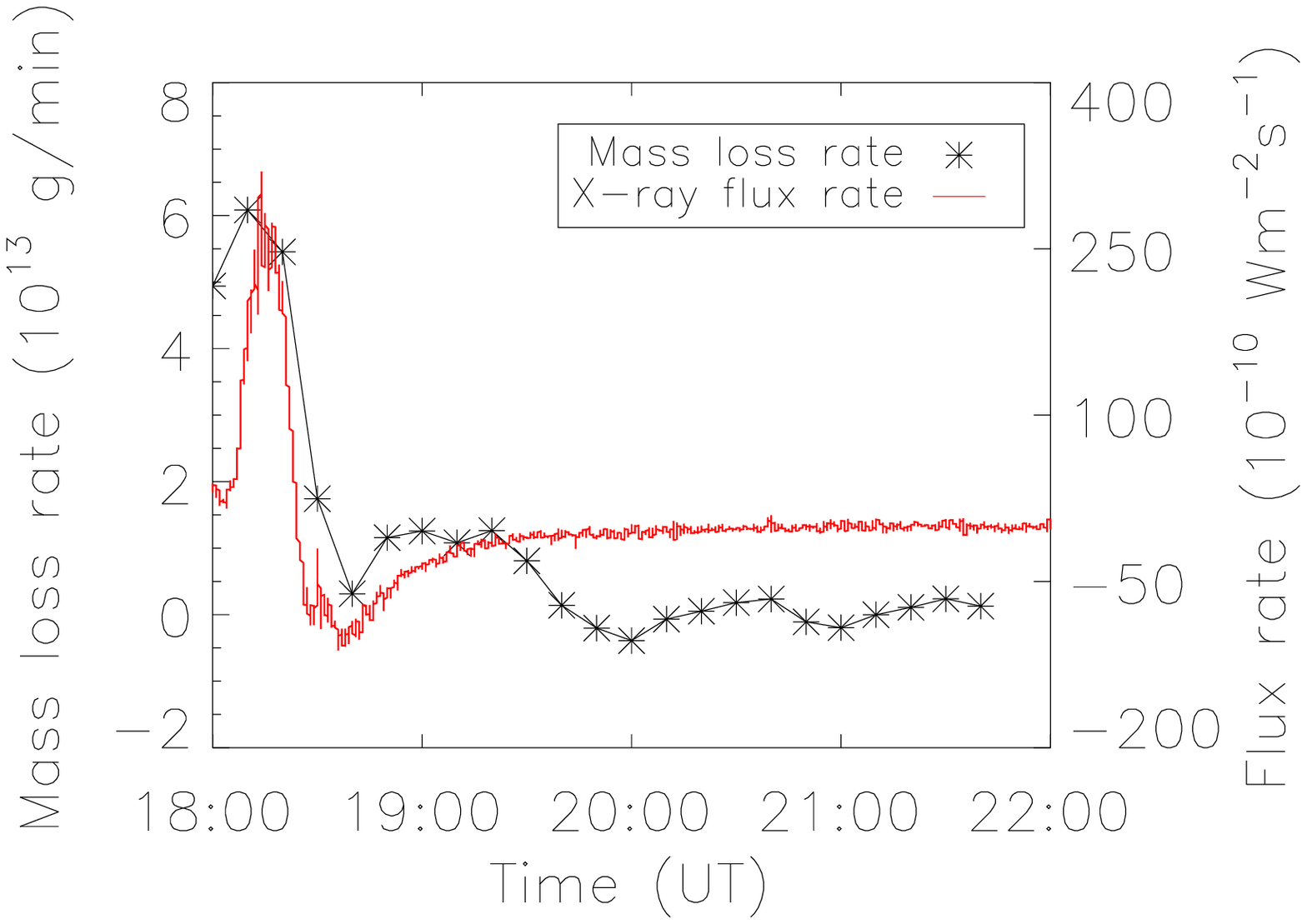}
              }
     \vspace{-0.36\textwidth}   % Shift close to the panel top 
     \centerline{\large \bf     % Includes the labels (here needs the color package)
      \hspace{0.06\textwidth} \color{black}{(c)}
      \hspace{0.45\textwidth}  \color{black}{(d)}
         \hfill}
     \vspace{0.36\textwidth}    % Shift back to the panel bottom 
%------------------------     

\caption{Left column: X-ray flux from GOES measurements (red solid line) and mass loss from the low corona (asterisks), for events 1 (a) and 2 (c). Right column: derivative of the X-ray flux (flux rate, red solid line) and of the mass loss (mass loss rate, asterisks).}
   \label{Fig5}
   \end{figure}

The fact that two of the analyzed events, namely events 1 and 2, are associated to X-ray flares, leads us to seek for temporal correspondences between the profile of the EUV mass evacuation and that of the respective flares described by the variation of the X-ray flux. Event 3 is not associated to a catalogued flare, and from direct inspection of GOES data we could not detect any increase in the soft X-ray flux at the time of the eruption. In Figure \ref{Fig5} we plot together the EUV mass loss evolution and the X-ray flux profile registered by GOES (0.1\,--\,0.8\,nm) for events 1 and 2 (panels a and c). For Event 1, Figure \ref{Fig5}a reveals evacuation of mass from the low corona minutes before the X-ray flux increment. This time difference could be either due to plasma flows before the eruption \citep{Vourlidas2012}, or due to the initial launch phase of the CME. For Event 2, the first EUV mass estimation available after the dimming’s onset matches the beginning of the flare’s rising phase. It is worth noting that this event is very impulsive, associated with an M-class flare; as opposite to Event 1, which is related to a slower-rising B-class flare. 

Figures \ref{Fig5}b and d display for events 1 and 2 the derivative of the X-ray flux in comparison with that of the mass loss, as a function of time. The derivative profiles in Figure \ref{Fig5}b exhibit a similar slope after the flare's onset and even after the evacuated mass reaches its maximum value. As for Event 2 (Figure \ref{Fig5}d) a good temporal agreement between the flare's peak and the time of fastest mass evacuation is observed. Several authors have previously reported a close correspondence between soft X-ray flux and acceleration profiles of CMEs \citep[\eg][]{Zhang2004, Vourlidas2012}. Since the dimmings studied in this work are associated with evacuation of plasma, which implies acceleration of material during the eruption, a good concordance between the mass loss and X-ray flux profiles is reasonable to be expected.

\section{Summary and Conclusions}
\label{S-Summary}

We report on the analysis of three coronal dimmings and their associated CMEs, whereby we apply separate techniques to estimate both the CME mass and the evacuated low coronal mass from the associated EUV dimmings, and analyze their relation.

The evacuated coronal mass is estimated with a DEM technique applied to the dimming regions using data from 3 AIA coronal passbands. The mass loss color maps deduced for each of the dimming regions indicate that the evacuation of plasma is not uniform. Mass loss is highest in some of the periphery regions of the bright post-eruptive loops, with Event 1 exhibiting the double-dimming pattern at the location of the ejected filament footpoints \citep{Thompson2000}. 

We determined the mass loss during several hours after the events, whose temporal behavior is different in the three cases. Although a stage of similar high evacuation rate is common for the three events, the following phase differs, with one of them presenting a gradual recovery (Event 1), another showing further small evacuation (Event 3), and the other exhibiting none of them, which could also imply that both recovery and evacuation compensate each other (Event 2). These three behaviors can be described with the aid of an exponential fit followed by a linear one.

Height-time measurements of the leading edge of the CMEs from the low corona and up to the reach of the STEREO coronagraphs indicate that the highest evacuation of mass takes place when the CMEs are traveling through the FOV of EUVI and COR1. Even when the CMEs are developed in the FOV of COR2, mass is still being evacuated from the low corona, though at a considerable lower rate.

The temporal evolution of dimming and CME AWs are in good agreement with each other, and also in good correspondence with the respective mass loss profiles. The similarity between the AW and the mass loss profiles is remarkable, suggesting that both attributes are closely related. The analysis of the AW behavior in time suggests that most of the expansion takes place at higher altitudes for the events associated with larger flares, at least for the three events here investigated. 

For the two events associated to flares, we examine the correspondence between the soft X-ray flux registered by GOES and the mass loss profiles. The beginning of mass evacuation precedes the flare's rising phase in Event 1, while both concur in Event 2. In addition, there is a good match between the peaks of both evacuated mass and X-ray flux derivatives. 
  
The temporal evolution of the dimming area, the CME AW, and the flare X-ray flux associated to each low-coronal event suggest a close connection to the process of mass evacuation in the EUV low corona.

The mass of the associated white-light CMEs was determined using COR2 total-brightness images. The total mass loss determined from EUV data differs by $\approx$\,33\,--\,63\,\% with respect to the white-light CME mass, depending of the event. Still, the evacuated mass from the coronal dimmings represents a considerable amount of the mass of the associated CMEs. Our results using AIA data represent a similar fraction of the CME mass with regards to other reports based on data from other instruments \citep[\eg][]{Harrison2003,Aschwanden2009,Tian2012}. At the same time, as opposed to some of these reported results, we find that the low-coronal evacuated mass determined from a DEM analysis is not sufficient to account for all the mass measurable in the white-light associated CMEs. 
Several aspects play a key role in the accuracy of evacuated mass and white-light mass determination. The evacuated mass in the low corona is certainly underestimated and should be taken as a lower limit due to the following:

\begin{itemize}
\item Prominences represent the brightest part of CMEs in white-light coronagraphic images, forming the bright core of these events. Their mass measured in EUV is in the range 10$^{14}$\,--\,10$^{15}$\,g \citep{Gilbert2006}, providing a considerable contribution to the mass of CMEs. This material is supposed to be of chromospheric origin and thus at a lower temperature. Therefore the characteristic temperature of the EUV passbands here analyzed may not account for the contribution of mass of this cooler material.
\item At the same time, prominence material on the disk is seen in absorption in EUV images, decreasing the emission measure obtained at the pre-event time while bright post-eruptive loops cover a significant area of the eruption site, where most of the evacuation is supposed to take place. 
\item In addition, because of the temperature covered by the used AIA passbands, the contribution of hot plasma ($\geq$\,2.5\,MK) is not considered in the determination of the evacuated mass. 
\end{itemize}

\noindent On the other hand, the difference with the mass of the CME estimated from white-light observations, can be at least partly explained by  the reasons below:

\begin{itemize}
\item The CME mass is supposedly composed not only of plasma ejected from the low corona, but also of mass provided by the cold prominence and by material being piled-up along the CME propagation \citep{Feng2015}. 
\item The difference between the white-light CME mass and the EUV evacuated mass, can also be attributed to the limitations in the methods used to determine the mass from white-light observations. For instance, the strong dependence on the relative position of the observer with respect to the direction of propagation, and the unknown density distribution within CMEs \citep{Vourlidas2000}, involve significant uncertainties in the CME mass determination. 
\end{itemize}

We are currently performing similar analyses on a larger sample of events. We intend to find associations among parameters related to EUV mass loss and white-light CME mass that may be helpful to deduce CME properties from low coronal data. Furthermore,  we pursue to gain understanding on how the sources of errors affect events with different characteristics.

\begin{acks}

FML and FAN are fellows of CONICET. HC and AMV are members of the Carrera del Investigador Cient\'ifico (CONICET). FML, HC, and LAB appreciate support from project UTN UTI1744. The authors thank the anonymous referee for very constructive comments and suggestions, as well as Marc L. DeRosa for helpful discussions. The authors acknowledge the use of data from the SDO/AIA and STEREO/SECCHI projects. The AIA data used here are courtesy of SDO (NASA) and the AIA consortium. The SECCHI data are courtesy of STEREO and the SECCHI consortium.

\end{acks}

\bibliographystyle{spr-mp-sola}
\bibliography{references_lopez}

\end{article} 

\end{document}